\documentclass[pre,aps,showpacs,byrevtex,onecolumn]{revtex4}
\usepackage{graphicx}
\usepackage{amsmath}

\begin{document}

\title{Semiclassical Diagonalization of Quantum Hamiltonian and Equations of Motion
with Berry Phase Corrections}
\author{Pierre Gosselin$^{1}$, Alain B\'{e}rard$^{2}$ and Herv\'{e} Mohrbach$^{2}$}

\address{$^1$Institut Fourier, UMR
5582 CNRS-UJF, UFR de Math\'ematiques, Universit\'e Grenoble I,
BP74, 38402 Saint Martin
d'H\`eres, Cedex, France \\
$^2$ Laboratoire de Physique Mol\'eculaire et des Collisions,
ICPMB-FR CNRS 2843, \\ Universit\'e Paul Verlaine-Metz,   57078
Metz Cedex 3, France}

\begin{abstract}
It has been recently found that the equations of motion of several
semiclassical systems must take into account terms arising from Berry phases
contributions. Those terms are responsible for the spin Hall effect in
semiconductor as well as the Magnus effect of light propagating in
inhomogeneous media. Intensive ongoing research on this subject seems to
indicate that a broad class of quantum systems may be affected by Berry
phase terms. It is therefore important to find a general procedure allowing
for the determination of semiclassical Hamiltonian with Berry Phase
corrections. This article presents a general diagonalization method at order
$\hbar $ for a large class of quantum Hamiltonians directly inducing Berry
phase corrections. As a consequence, Berry phase terms on both coordinates
and momentum operators naturally arise during the diagonalization procedure.
This leads to new equations of motion for a wide class of semiclassical
system. As physical applications we consider here a Dirac particle in an
electromagnetic or static gravitational field, and the propagation of a
Bloch electrons in an external electromagnetic field.
\end{abstract}

\maketitle

\section{Introduction}

Since the seminal work of Berry \cite{BERRY}, the notion of Berry phase has
found several applications in branches of quantum physics such as atomic and
molecular physics, optics and gauge theories. Most studies consider the
geometric phase that a wave function acquires when a quantum mechanical
system has an adiabatic evolution. Yet, the Berry phase in momentum space
has recently found unexpected applications in the topic of spintronics. Such
a term may indeed be responsible for a transverse dissipationless
spin-current in semiconductors in the presence of electric fields \cite
{MURAKAMI1}. This effect is a particular case of the Spin-Hall effect which
is now predicted and observed in many different physical situations and can
be interpreted at the semiclassical level as due to the influence of Berry
connections on semiclassical equations of motion of spinning particles, like
electrons in electric \cite{ALAIN} or magnetic fields \cite{BLIOKH1}. In the
above cited examples, the semiclassical equations of motion where derived
from an approximate semiclassical Hamiltonian in a representation where this
latter is diagonal. It was then shown that a noncommutative geometry,
originating from the presence of a Berry phase which turns out to be a
spin-orbit coupling, underlies the semiclassical dynamics. Spin-orbit
contributions on the propagation of light have also been the focus of
several other works \cite{ALAIN,BLIOKH2, MURAKAMI2} and have led to a
generalization of geometric optics called geometric spinoptics \cite
{HORVATHY1}.

Semiclassical methods play a very important role in solid state physics too,
in studying the dynamics of electrons to account for the various properties
of metals, semiconductors and insulators \cite{MERMIN}. In a series of
papers \cite{NIU1} (see also \cite{SHINDOU}), a new set of semiclassical
equations with a Berry phase correction was proposed to account for the
semiclassical dynamics of electrons in magnetic Bloch bands (in the usual
one band approximation). These equations were derived by considering a wave
packet in a band and using a time-dependent variational principle in a
Lagrangian formulation. The derivation of a semiclassical Hamiltonian was
shown to lead to difficulties in the presence of Berry phase terms \cite
{NIU1}. The apparent non-canonical character of the equations of motion with
Berry phase corrections led the authors of \cite{NIU2} to conclude that the
naive phase space volume is not conserved in the presence of a Berry phase
and a magnetic field. There is nevertheless an invariant measure as found by
themselves, so the Liouville theorem is not really violated in the end. This
invariant measure is actually a general result of the well-established
theory of non-canonical Hamiltonian dynamics, as pointed out by a number of
authors \cite{HORVATHY3,BLIOKH3,GHOSH}. However it is only in \cite{PIERRE}
that the non-canonical Hamiltonian formulation of a semiclassical electron
in magnetic Bloch bands has been fully derived. This Hamiltonian approach
allows deriving rigorously the semi classical equations of motion, including
explicitly the role of the Berry curvature and showed many similarities with
the description of a Dirac electron in an electromagnetic field as in \cite
{BLIOKH1}. The common feature of the Hamiltonian formulations discussed
above is that a noncommutative geometry underlies the algebraic structure of
both coordinates and momenta. Actually, a Berry phase contribution to the
dynamical operators stems from the representation where the kinetic energy
is diagonal (for instance Foldy-Wouthuysen or Bloch representation). In this
representation the physical coordinate and momentum become noncommutative
operators.

The previous discussion shows that Berry phase terms could be present in
semiclassical equations of motion of several physical systems ranging from
electrons in vacuum, in solid or in semiconductor to photons in
inhomogeneous media, with potential application in the field of spintronics
and spinoptics. This in turn called for a general semiclassical Hamiltonian
formalism from which semiclassical dynamics of a quantum system can be
derived. This paper presents a general method of diagonalization at order $%
\hbar $ for a quantum mechanical Hamiltonian presenting bands structure,
like for electrons in a periodic potential or for a Dirac (massive or
massless) like-Hamiltonian. Starting with a Hamiltonian depending only on an
invariant momentum $\mathbf{P}$ and whose diagonalization is known, we can
introduce a dependence in the variable $\mathbf{R}$ and diagonalize the
Hamiltonian in four steps (discussed in the text). During this process of
diagonalization, we show that both position and momentum operators acquire a
Berry-phase contribution making both the coordinate and momentum algebra
noncommutative. As physical applications and to check to validity of our
method we further consider the case of Dirac particle in an electromagnetic
field and compare with the semiclassical diagonalization given in \cite
{BLIOKH1}. We also consider the case of a Dirac particle in a symmetric
static gravitational field (the asymmetric case is studied in ref. \cite
{PHOTONPIERRE}) and compare with the articles \cite{OBUKHOV,SILENKO}. As a
last application, the reader will find the details of the diagonalization
sketched in \cite{PIERRE}.for the propagation of a Bloch electron (spinless)
in an external electromagnetic field. These various physical applications
show that our semiclassical Hamiltonian diagonalization approach is
potentially promising since it unifies several apparently unrelated
problematics in one formalism.

The paper is organized as follows. In section II we develop our formalism in
the case of a general Hamiltonian which has an energy bands structure. We
then derive the very general equations of motion in this case. Section III
is devoted to the application of our method to the case of the Dirac
Hamiltonian in an electromagnetic field but in a flat space, and to the
diagonalization of the Dirac Hamiltonian in a symmetric static gravitational
field, allowing us to check the validity of our method. Section IV retrieves
the equations of motion for an electron in a periodic potential within our
general set up. Section V is for the conclusion.

\section{ A general process of semiclassical diagonalization.}

\textbf{I}n this section we present a method to diagonalize at the
semiclassical order ($\hbar $) a quantum mechanical system whose state space
is a tensor product $L^{2}\left( \mathcal{R}^{3}\right) \otimes V$\ with $V$%
\ some internal space. In other words, the Hamiltonian of this system can be
written as a matrix $H_{0}\left( \mathbf{P,R}\right) $\ of size $\dim V$\
whose elements are operators depending on a couple of canonical variables $P$%
\ and $R$. The archetype example is the Dirac Hamiltonian with $V=C^{4}$,
but in appendix $2$\ we show how a spinless electron in a periodic potential
fits also in this set up. By diagonalization, we always mean here a unitary
transformation setting the Hamiltonian in a diagonal matrix form, the
diagonal elements being operators depending on $P$\ and $R$. That is, we do
not aim at finding the eigenvalues, but rather to derive the Band
Hamiltonians, that are usually relevant for the semiclassical dynamics.

More precisely, our strategy will be to solve the exact diagonalization for $%
H_{0}\left( \mathbf{P,R}\right) $\ at order $\hbar $, when the
diagonalization of a fictitious Hamiltonian $H_{0}\left( \mathbf{P,}%
\widetilde{\mathbf{r}}\right) $\ is known for a parameter $\widetilde{%
\mathbf{r}}$\ (replacing $R$) which is supposed to commute with $P$\ (for
instance this may be the Foldy-Wouthuysen transformation \cite{FW} for a
free Dirac particle). We show how to compute the quantum corrections (at
leading order in $\hbar $) that were neglected during this formal
diagonalization (where position and momenta where considered as commuting
quantities). The idea behind this procedure is that it is much easier to
solve the diagonalization for $H_{0}\left( \mathbf{P,}\widetilde{\mathbf{r}}%
\right) $, as seen in our applications, and only then turn to $H_{0}\left(
\mathbf{P,R}\right) $.

\subsection{ Preliminary : Products of operators series.}

To develop our process of diagonalization, the semiclassical expression of
products of symmetrized expressions $S\left( \mathbf{P},\mathbf{R}\right) $
depending on the canonical couple of variables $\mathbf{P}$ and $\mathbf{R}$
is required. These expressions are also assumed to have series expansions in
$\mathbf{P}$ and $\mathbf{R}$ whose coefficients can be of a matrix form
(this last assumption allowing to deal with Dirac Hamiltonians). Let us
consider two such expressions $S_{1}\left( \mathbf{P},\mathbf{R}\right) $
and $S_{2}\left( \mathbf{P},\mathbf{R}\right) $, supposed to be symmetrized
in $\mathbf{P}$ and $\mathbf{R}$. By symmetrization, we mean that each
expression has been written in a form where all the powers of $\mathbf{P}$
have been put half on the left and half on the right of the expression. Our
aim is now to write the product $S_{1}\left( \mathbf{P},\mathbf{R}\right)
S_{2}\left( \mathbf{P},\mathbf{R}\right) $ as a symmetric expression in
terms of $\mathbf{P}$ and $\mathbf{R}$. This is easy to realize at order $%
\hbar $, since in that case, pushing half of the powers of $\mathbf{P}$ in $%
S_{2}\left( \mathbf{P},\mathbf{R}\right) $ on the left and half the powers
of $S_{1}\left( \mathbf{P},\mathbf{R}\right) $ on the right is equivalent to
the computation of some commutators. One can easily see that at order $\hbar
$
\begin{equation}
S_{1}\left( \mathbf{P},\mathbf{R}\right) S_{2}\left( \mathbf{P},\mathbf{R}%
\right) =Sym(S_{1}\left( \mathbf{P},\mathbf{R}\right) S_{2}\left( \mathbf{P},%
\mathbf{R}\right) )+\frac{i}{2}\hbar Asym\nabla _{R_{l}}\nabla _{P^{l}}\left[
S_{1}\left( \mathbf{P},\mathbf{R}\right) S_{2}\left( \mathbf{P},\mathbf{R}%
\right) \right]  \label{SS}
\end{equation}
where $Sym(S_{1}\left( \mathbf{P},\mathbf{R}\right) S_{2}\left( \mathbf{P},%
\mathbf{R}\right) )$ is the symmetrized expression for the product, and $%
Asym $ is defined through
\begin{equation}
Asym\nabla _{R_{l}}\nabla _{P^{l}}\left[ S_{1}\left( \mathbf{P},\mathbf{R}%
\right) S_{2}\left( \mathbf{P},\mathbf{R}\right) \right] =\nabla
_{R_{l}}S_{1}\left( \mathbf{P},\mathbf{R}\right) \nabla _{P^{l}}S_{2}\left(
\mathbf{P},\mathbf{R}\right) -\nabla _{P^{l}}S_{1}\left( \mathbf{P},\mathbf{R%
}\right) \nabla _{R_{l}}S_{2}\left( \mathbf{P},\mathbf{R}\right)
\end{equation}
This formula can be easily generalized to an arbitrary product of $n$ terms,
but for the sequel of this paper, only three will be needed
\begin{eqnarray}
S_{1}\left( \mathbf{P},\mathbf{R}\right) S_{2}\left( \mathbf{P},\mathbf{R}%
\right) S_{3}\left( \mathbf{P},\mathbf{R}\right) &=&Sym(S_{1}\left( \mathbf{P%
},\mathbf{R}\right) S_{2}\left( \mathbf{P},\mathbf{R}\right) S_{3}\left(
\mathbf{P},\mathbf{R}\right) )  \notag \\
&&+\frac{i}{2}\hbar Asym\nabla _{R_{l}}\nabla _{P^{l}}\left[ S_{1}\left(
\mathbf{P},\mathbf{R}\right) S_{2}\left( \mathbf{P},\mathbf{R}\right) \right]
S_{3}\left( \mathbf{P},\mathbf{R}\right)  \notag \\
&&+\frac{i}{2}\hbar S_{1}\left( \mathbf{P},\mathbf{R}\right) Asym\nabla
_{R_{l}}\nabla _{P^{l}}\left[ S_{2}\left( \mathbf{P},\mathbf{R}\right)
S_{3}\left( \mathbf{P},\mathbf{R}\right) \right]  \notag \\
&&+\frac{i}{2}\hbar Asym\nabla _{R_{l}}\nabla _{P^{l}}\left[ S_{1}\left(
\mathbf{P},\mathbf{R}\right) _{S_{2}\left( \mathbf{P},\mathbf{R}\right)
}S_{3}\left( \mathbf{P},\mathbf{R}\right) \right]  \label{SSS}
\end{eqnarray}
where
\begin{eqnarray}
&&Asym\nabla _{R_{l}}\nabla _{P^{l}}\left[ S_{1}\left( \mathbf{P},\mathbf{R}%
\right) _{S_{2}\left( \mathbf{P},\mathbf{R}\right) }S_{3}\left( \mathbf{P},%
\mathbf{R}\right) \right]  \notag \\
&=&\left[ \nabla _{R_{l}}S_{1}\left( \mathbf{P},\mathbf{R}\right) \right]
S_{2}\left( \mathbf{P},\mathbf{R}\right) \nabla _{P^{l}}S_{3}\left( \mathbf{P%
},\mathbf{R}\right) -\left[ \nabla _{P^{l}}S_{1}\left( \mathbf{P},\mathbf{R}%
\right) \right] S_{2}\left( \mathbf{P},\mathbf{R}\right) \nabla
_{R_{l}}S_{3}\left( \mathbf{P},\mathbf{R}\right)
\end{eqnarray}
Let us stress again that all this identities are valid only at order $\hbar $
and that considering higher order corrections would of course induce more
corrections.

\subsection{ Diagonalization with a parameter $\widetilde{r}$}

Let us consider a general Hamiltonian $H_{0}\left( \mathbf{P,R}\right) $
which admits a series expansion in $\mathbf{P}$ and $\mathbf{R}$ written
here for convenience in a symmetrical form. To perform the semi classical
diagonalization of this operator, we first consider in this subsection a
fictitious Hamiltonian $H_{0}\left( \mathbf{P,}\widetilde{\mathbf{r}}\right)
$ where a parameter $\widetilde{\mathbf{r}}$ commuting with $\mathbf{P}$ has
replaced the operator $\mathbf{R}$. We further assume that $H_{0}\left(
\mathbf{P,}\widetilde{\mathbf{r}}\right) $ is known to be exactly
diagonalized through a matrix $U\left( \mathbf{P,}\widetilde{\mathbf{r}}%
\right) \equiv U$. We assume that $U$\ and $U\left( \mathbf{P,}\widetilde{%
\mathbf{r}}\right) H_{0}\left( \mathbf{P,}\widetilde{\mathbf{r}}\right)
U^{+}\left( \mathbf{P,}\widetilde{\mathbf{r}}\right) $\ can be expanded in
series of $P$ (with positive or negative powers)\ as it will be the case in
our applications, but in fact this assumption which is practical for our
proofs could probably be relaxed.

As an typical example we can consider the following kind of Dirac
Hamiltonian
\begin{equation}
H_{0}=\mathbf{\alpha .}\left( \mathbf{P-A(}\widetilde{\mathbf{r}}\mathbf{)}%
\right) \mathbf{+}\beta m
\end{equation}
where $\mathbf{A(}\widetilde{\mathbf{r}}\mathbf{)}$ mimics a formal magnetic
potential. The potential $\mathbf{A(}\widetilde{\mathbf{r}}\mathbf{)}$ being
$\mathbf{R}$ independent, it only shifts the momentum for each value of $%
\widetilde{\mathbf{r}}$. As a consequence, the usual Foldy Wouthuysen \cite
{FW} transformation expressed in terms of $\mathbf{P-A}$ instead of $\mathbf{%
P}$, diagonalizes the Dirac Hamiltonian exactly.

Going back to the general case, the diagonalization of $H_{0}\left( \mathbf{%
\ P,}\widetilde{\mathbf{r}}\right) $ will be written as
\begin{equation}
\varepsilon \left( \mathbf{P,}\widetilde{\mathbf{r}}\right) =U\left( \mathbf{%
P,}\widetilde{\mathbf{r}}\right) H_{0}\left( \mathbf{P,}\widetilde{\mathbf{r}%
}\right) U^{+}\left( \mathbf{P,}\widetilde{\mathbf{r}}\right)
\end{equation}
where $\varepsilon \left( \mathbf{P,}\widetilde{\mathbf{r}}\right) $ is a
diagonal matrix. For later use, let us notice that we have $U\left( \mathbf{%
P,}\widetilde{\mathbf{r}}\right) \mathbf{P}U^{+}\left( \mathbf{P,}\widetilde{%
\mathbf{r}}\right) =\mathbf{P}$.

To gain some hints from our initial diagonalization, recall we assumed that
the product $\varepsilon \left( \mathbf{P,}\widetilde{\mathbf{r}}\right)
=U\left( \mathbf{P,}\widetilde{\mathbf{r}}\right) H_{0}\left( \mathbf{P,}%
\widetilde{\mathbf{r}}\right) U^{+}\left( \mathbf{P,}\widetilde{\mathbf{r}}%
\right) $ can be expanded in series of monomial terms of the form : $\left(
\varepsilon _{1}^{i_{1}}\mathbf{(}\widetilde{\mathbf{r}}\mathbf{)}%
.P_{i_{1}}^{k_{i_{1}}}\right) ...\left( \varepsilon _{n}^{i_{n}}\mathbf{(}%
\widetilde{\mathbf{r}}\mathbf{)}.P_{i_{n}}^{k_{i_{n}}}\right) $ ($\ast $)
where the $\varepsilon _{l}^{i_{l}}\mathbf{(}\widetilde{\mathbf{r}}\mathbf{)}
$ are some matrices commuting with $\mathbf{P}$, and $P_{i_{l}}^{k_{i_{l}}}$
is the $i_{l}$-th component of $\mathbf{P}$ ($i_{l}=1,2,3)$ at some power $%
k_{i_{l}}$ (this power should be written $k_{i_{l},l}$ but we will avoid an
excess of notation here). As usual, the sums over the $i_{l}$ are implicit.
Rearranging the series in powers of $P_{1}$, $P_{2}$, $P_{3}$ and given that
$\widetilde{\mathbf{r}}$ is a parameter, we can write the energy in a
symmetrical form :
\begin{equation}
\varepsilon \left( \mathbf{P,}\widetilde{\mathbf{r}}\right) =\frac{1}{2}%
\sum_{\mathbf{X}}\left[ A_{\mathbf{X}}\mathbf{(}\widetilde{\mathbf{r}}%
\mathbf{)}\left( \prod_{i=1,2,3}P_{i}^{X_{i}}\right) +\left(
\prod_{i=1,2,3}P_{i}^{X_{i}}\right) A_{\mathbf{X}}\mathbf{(}\widetilde{%
\mathbf{r}}\mathbf{)}\right]
\end{equation}
where $\left( \prod_{i=1,2,3}P_{i}^{X_{i}}\right) $ is a given momentum
power and $A_{\mathbf{X}}\mathbf{(}\widetilde{\mathbf{r}}\mathbf{)}$ is a
combination of the $\varepsilon _{l}^{i_{l}}\mathbf{(}\widetilde{\mathbf{r}}%
\mathbf{)}$. The $\mathbf{X}$ labels the multi index $\left( X_{i},\text{ }%
i=1,2,3\right) $.

The important consequence here is that the matrix \ $A_{\mathbf{X}}\mathbf{(}%
\widetilde{\mathbf{r}}\mathbf{)}$ is diagonal.\ As an example, think of the
Dirac Hamiltonian diagonalization, which involves some products $\mathbf{%
\alpha }.\mathbf{P\alpha }.\mathbf{P}$, rearranged as
\begin{equation}
\frac{1}{2}\left[ \alpha _{i}P_{i}\alpha _{j}P_{j}+\alpha _{j}P_{j}\alpha
_{i}P_{i}\right] =\frac{1}{2}\left[ \alpha _{i}\alpha _{j}+\alpha _{j}\alpha
_{i}\right] P_{i}P_{j}
\end{equation}
and $\frac{1}{2}\left[ \alpha _{i}\alpha _{j}+\alpha _{j}\alpha _{i}\right] $
is diagonal, by the usual rules for Dirac matrices.

Let us conclude this subsection by noting that the symmetrizations we
performed is unnecessary here, but will be of a practical interest when
dealing with the exact diagonalization.

\subsection{ Introducing the $R$ dependence. The transformation Ansatz.}

We are now going to reintroduce $\mathbf{R}$ into $H_{0}$ in order to
diagonalize exactly $H_{0}\left( \mathbf{P,R}\right) $ at the $\hbar $
order. In the following symmetrization in $\mathbf{R}$ and $\mathbf{P}$ will
be assumed in all expressions. To find the diagonalization transformation
for $H_{0}\left( \mathbf{P,R}\right) $, we will use the following method.
First notice that the Hamiltonian $H_{0}\left( \mathbf{P,R}\right) $ is
''almost'' diagonalized, that is diagonalized at zeroth order in $\hbar $,
through the transformation $U\left( \mathbf{P},\mathbf{R}\right) $ (which is
also symmetrized given our convention).

However, through the symmetrization process, the matrix $U\left( \mathbf{P},%
\mathbf{R}\right) $ does not remain unitary. As a consequence, we will
rather consider a matrix $U\left( \mathbf{P},\mathbf{R}\right) +XU\left(
\mathbf{P},\mathbf{R}\right) $, where $X$ is a contribution of order $\hbar $
that ensures the unitarity of the transformation at order $\hbar $ (the
factor $U$ is a normalization that simplifies the subsequent expressions).

The matrix $X$ can be explicitly computed. Actually, from the unitary
conditions
\begin{equation}
\left( U(\mathbf{P},\mathbf{R)}+XU\right) \left( U^{+}(\mathbf{P},\mathbf{R)}%
+U^{+}X^{+}\right) =1
\end{equation}
and
\begin{equation}
\left( U^{+}(\mathbf{P},\mathbf{R)}+U^{+}X^{+}\right) \left( U(\mathbf{P},%
\mathbf{R)}+XU\right) =1
\end{equation}
or equivalently
\begin{eqnarray}
U(\mathbf{P},\mathbf{R)}U^{+}(\mathbf{P},\mathbf{R)}+X+X^{+} &=&1  \notag \\
U^{+}(\mathbf{P},\mathbf{R)}U(\mathbf{P},\mathbf{R)}+U^{+}\left(
X+X^{+}\right) U &=&1
\end{eqnarray}
To solve this equation, let us first notice that $U(\mathbf{P},\mathbf{R)}%
U^{+}(\mathbf{P},\mathbf{R)}\neq 1$ since $U(\mathbf{P},\mathbf{R)}$ is not
unitary. The crucial point here and in the sequel of this paper is the
computation of a product of expressions in which the $\mathbf{R}$ dependence
has been introduced. To do so let us use the initial relation for the
parameter $\widetilde{\mathbf{r}}$
\begin{equation}
U(\mathbf{P},\widetilde{\mathbf{r}}\mathbf{)}U^{+}(\mathbf{P},\widetilde{%
\mathbf{r}}\mathbf{)=}1
\end{equation}
expanded in a symmetric series (as we did for the Hamiltonian)
\begin{equation}
U(\mathbf{P},\widetilde{\mathbf{r}}\mathbf{)}U^{+}(\mathbf{P},\widetilde{%
\mathbf{r}}\mathbf{)}=\frac{1}{2}\sum_{\mathbf{X}}\left[ B_{\mathbf{X}}%
\mathbf{(}\widetilde{\mathbf{r}}\mathbf{)}\left(
\prod_{i=1,2,3}P_{i}^{X_{i}}\right) +\left(
\prod_{i=1,2,3}P_{i}^{X_{i}}\right) B_{\mathbf{X}}\mathbf{(}\widetilde{%
\mathbf{r}}\mathbf{)}\right] =1.
\end{equation}
Therefore the series expansion $\sum_{\mathbf{X}}B_{\mathbf{X}}\mathbf{(}%
\widetilde{\mathbf{r}}\mathbf{)}\left( \prod_{i=1,2,3}P_{i}^{X_{i}}\right) $
reduces to one constant term : the identity matrix.

Now, going back to $U(\mathbf{P},\mathbf{R)}$, we use the symmetrization
formula (\ref{SS})
\begin{eqnarray}
U(\mathbf{P},\mathbf{R)}U^{+}(\mathbf{P},\mathbf{R)} &=&\frac{1}{2}\sum_{%
\mathbf{X}}\left[ B_{\mathbf{X}}\mathbf{(R)}\left(
\prod_{i=1,2,3}P_{i}^{X_{i}}\right) +\left(
\prod_{i=1,2,3}P_{i}^{X_{i}}\right) B_{\mathbf{X}}\mathbf{(R)}\right]  \notag
\\
&&+\frac{i}{2}\hbar Asym\nabla _{R_{l}}\nabla _{P^{l}}\left[ U(\mathbf{P},%
\mathbf{R)}U^{+}(\mathbf{P},\mathbf{R)}\right]
\end{eqnarray}
Since our result about the $B_{\mathbf{X}}\mathbf{(}\widetilde{\mathbf{r}}%
\mathbf{)}$ applies also to $B_{\mathbf{X}}\mathbf{(R)}$ after replacing $%
\widetilde{\mathbf{r}}$ by $\mathbf{R}$, one has :
\begin{eqnarray}
U(\mathbf{P},\mathbf{R)}U^{+}(\mathbf{P},\mathbf{R)} &=&1+\frac{i}{2}\hbar
Asym\nabla _{R_{l}}\nabla _{P^{l}}\left[ U(\mathbf{P},\mathbf{R)}U^{+}(%
\mathbf{P},\mathbf{R)}\right]  \notag \\
&=&1-\frac{i}{2\hbar }\left[ \mathcal{A}_{P_{l}},\mathcal{A}_{R_{l}}\right]
\end{eqnarray}
where we have defined the (''non projected'', see below) Berry connections
(we use the terminology of ref.\cite{BLIOKH4}) as
\begin{eqnarray}
\mathcal{A}_{\mathbf{R}} &=&i\hbar U(\mathbf{P},\mathbf{R)}\nabla _{\mathbf{P%
}}U^{+}(\mathbf{P},\mathbf{R)}  \notag \\
\mathcal{A}_{\mathbf{P}} &=&-i\hbar U(\mathbf{P},\mathbf{R)}\nabla _{\mathbf{%
R}}U^{+}(\mathbf{P},\mathbf{R)}.
\end{eqnarray}
that are non diagonal Hermitian matrices of order $\hbar $. Similarly, we
have also:
\begin{eqnarray}
U^{+}(\mathbf{P},\mathbf{R)}U(\mathbf{P},\mathbf{R)} &=&1+\frac{i}{2}\hbar
Asym\nabla _{R_{l}}\nabla _{P^{l}}\left[ U^{+}(\mathbf{P},\mathbf{R)}U(%
\mathbf{P},\mathbf{R)}\right]  \notag \\
&=&1-U^{+}(\mathbf{P},\mathbf{R)}\frac{i}{2\hbar }\left[ \mathcal{A}_{P^{l}},%
\mathcal{A}_{R_{l}}\right] U(\mathbf{P},\mathbf{R)}
\end{eqnarray}
Let us note that, as can be checked easily, the Berry phases are Hermitian.
So is $\frac{i}{2\hbar }\left[ \mathcal{A}_{P^{l}},\mathcal{A}_{R_{l}}\right]
$. Therefore we can solve our problem with
\begin{eqnarray}
X &=&\frac{i}{4\hbar }\left[ \mathcal{A}_{P^{l}},\mathcal{A}_{R_{l}}\right]
\notag \\
X^{+} &=&\frac{i}{4\hbar }\left[ \mathcal{A}_{P^{l}},\mathcal{A}_{R_{l}}%
\right]
\end{eqnarray}
Let us make an important remark at this point. Our choice for $X$ is
obviously not unique. Actually, it has been chosen to ensure the unitarity
of the transformation and to obtain a transformation that reduces to the
initial one when $\mathbf{P}$ and $\mathbf{R}$ do commute. We could thus add
to $X$ an expression like $\delta XU$ where $\delta X$ is anti-Hermitian. It
is easy to see that the operator $U\left( \mathbf{P},\mathbf{R}\right)
+XU+\delta XU$ is still unitary. However, we will soon see that this non
unicity is irrelevant and that our choice is sufficient to perform the
diagonalization at order $\hbar $.

\subsection{The quasidiagonalization}

We will now consider the following quasi-diagonalization transformation
\begin{equation}
\left[ \left( U\left( \mathbf{P},\mathbf{R}\right) +XU\right) H_{0}\left(
\mathbf{P,R}\right) \left( U^{+}\left( \mathbf{P,R}\right)
+U^{+}X^{+}\right) \right]
\end{equation}
To compute this last expression, decompose it at the first order in $\hbar $
as
\begin{eqnarray}
&&U\left( \mathbf{P},\mathbf{R}\right) H_{0}\left( \left( \mathbf{P,R}%
\right) \right) U^{+}\left( \mathbf{P,R}\right) +XUH_{0}\left( \mathbf{P,R}%
\right) U^{+}\left( \mathbf{P,R}\right) +U\left( \mathbf{P},\mathbf{R}%
\right) H_{0}\left( \mathbf{P,R}\right) U^{+}X^{+}  \notag \\
&\simeq &U\left( \left( \mathbf{P},\mathbf{R}\right) \right) H_{0}\left(
\left( \mathbf{P,R}\right) \right) U^{+}\left( \left( \mathbf{P,R}\right)
\right) +X\varepsilon \left( \left( \mathbf{P,R}\right) \right) +\varepsilon
\left( \mathbf{P,R}\right) X^{+}
\end{eqnarray}
Let us first have a look to $U\left( \mathbf{P},\mathbf{R}\right)
H_{0}\left( \mathbf{P,R}\right) U^{+}\left( \mathbf{P,R}\right) $ and
consider it, as before, as a series of products of operators in $\mathbf{P}$
and $\mathbf{R}.$If this two variables where commuting, we would recover the
expansion $\varepsilon \left( \mathbf{P,R}\right) $ given in the subsection
B. But now, since, $\mathbf{R}$ does not commute with $\mathbf{P}$, one has
rather
\begin{eqnarray}
U\left( \mathbf{P},\mathbf{R}\right) H_{0}\left( \mathbf{P,R}\right)
U^{+}\left( \mathbf{P,R}\right) &=&\frac{1}{2}\sum_{\mathbf{X}}\left[ A_{%
\mathbf{X}}\mathbf{(R)}\left( \prod_{i=1,2,3}P_{i}^{X_{i}}\right) +\left(
\prod_{i=1,2,3}P_{i}^{X_{i}}\right) A_{\mathbf{X}}\mathbf{(R)}\right]  \notag
\\
&&+\left[ \text{ commutators}\right]  \label{CC}
\end{eqnarray}
the commutators appearing while pushing the momentum powers on the left or
on the right.

We can compute the first two terms of the right hand side by the same trick
as before. Actually, by construction, the coefficients of the series
expansion of $A_{\mathbf{X}}\mathbf{(R)}$ in the variable $\mathbf{R}$, are
the same as the coefficients (which are diagonal) of the expansion of $A_{%
\mathbf{X}}\mathbf{(r)}$ in the parameter $\mathbf{r}$. As a consequence,
\begin{equation}
\frac{1}{2}\sum_{\mathbf{X}}\left[ A_{\mathbf{X}}\mathbf{(R)}\left(
\prod_{i=1,2,3}P_{i}^{X_{i}}\right) +\left(
\prod_{i=1,2,3}P_{i}^{X_{i}}\right) A_{\mathbf{X}}\mathbf{(R)}\right]
\end{equation}
is the series expansion of $\varepsilon \left( \mathbf{P,R}\right) $, the
powers of $\mathbf{P}$ being rejected symmetrically to the left and to the
right. \bigskip

As an example, consider again the case of the Dirac Hamiltonian with an
electromagnetic field. The free ''Benchmark'' case is $\varepsilon
^{2}\left( \mathbf{P}\right) =\mathbf{P}^{2}$, and given our conventions,
replacing $\mathbf{P}$ by $\mathbf{P-A}(\mathbf{R})$ leads us to define $%
\varepsilon ^{2}\left( \mathbf{P-A}(\mathbf{R})\right) =\mathbf{P}^{2}-%
\mathbf{A}(\mathbf{R}).\mathbf{P-P}.\mathbf{A}(\mathbf{R})+\mathbf{A}^{2}(%
\mathbf{R})$, which is simply the usual operator $\left( \mathbf{P-A}(%
\mathbf{R})\right) ^{2}$. By the same way, we obtain as a series expansion
\begin{equation}
\varepsilon \left( \mathbf{P-A}(\mathbf{R})\right) =\varepsilon \left(
\mathbf{P}\right) -\frac{1}{2}\left( \mathbf{A}(\mathbf{R}).\frac{\mathbf{P}%
}{\mathbf{P}^{2}}\mathbf{+}\frac{\mathbf{P}}{\mathbf{P}^{2}}.\mathbf{A}(%
\mathbf{R})\right) +\frac{1}{4}\left[ \frac{1}{\mathbf{P}^{2}}\mathbf{A}^{2}(%
\mathbf{R})+\mathbf{A}^{2}(\mathbf{R})\frac{1}{\mathbf{P}^{2}}\right] +...
\end{equation}
The last term in the right hand side of (\ref{CC}) involves just half the
commutators obtained in pushing the momentum operators to the left or the
right. As explained in the subsection A Eq.(\ref{SSS}), they are simply
given by
\begin{equation}
\left[ \text{commutators}\right] =\frac{i}{2}\hbar Asym\nabla _{R_{l}}\nabla
_{P^{l}}\left[ U\left( \mathbf{P},\mathbf{R}\right) H_{0}\left( \mathbf{P,R}%
\right) U^{+}\left( \mathbf{P,R}\right) \right]
\end{equation}
As a consequence, we can write
\begin{equation}
U\left( \mathbf{P},\mathbf{R}\right) H_{0}\left( \mathbf{P,R}\right)
U^{+}\left( \mathbf{P,R}\right) =\varepsilon \left( \mathbf{P,R}\right) +%
\frac{i}{2}\hbar Asym\nabla _{R_{l}}\nabla _{P^{l}}\left[ U\left( \mathbf{P},%
\mathbf{R}\right) H_{0}\left( \mathbf{P,R}\right) U^{+}\left( \mathbf{P,R}%
\right) \right]
\end{equation}
A lengthy but straightforward computation presented in appendix 1 leads to
\begin{eqnarray}
U\left( \mathbf{P},\mathbf{R}\right) H_{0}\left( \mathbf{P,R}\right)
U^{+}\left( \mathbf{P,R}\right) &=&\varepsilon \left( \mathbf{P,R}\right) +%
\frac{1}{2}\left[ \mathcal{A}_{R_{l}}\nabla _{R^{l}}\varepsilon \left(
\mathbf{P,R}\right) +\nabla _{R^{l}}\varepsilon \left( \mathbf{P,R}\right)
\mathcal{A}_{R_{l}}\right]  \notag \\
&&+\frac{1}{2}\left[ \mathcal{A}_{P^{l}}\nabla _{P_{l}}\varepsilon \left(
\mathbf{P,R}\right) +\nabla _{P_{l}}\varepsilon \left( \mathbf{P,R}\right)
\mathcal{A}_{P^{l}}\right]  \notag \\
&&-\frac{i}{2\hbar }\left[ \varepsilon \left( \mathbf{P,R}\right) ,\mathcal{A%
}_{P^{l}}\right] \mathcal{A}_{R_{l}}+\frac{i}{2\hbar }\left[ \varepsilon
\left( \mathbf{P,R}\right) ,\mathcal{A}_{R_{l}}\right] \mathcal{A}_{P^{l}}
\notag \\
&&+\frac{i}{2\hbar }\left[ \mathcal{A}_{R_{l}},\mathcal{A}_{P^{l}}\right]
\varepsilon \left( \mathbf{P,R}\right) ,
\end{eqnarray}
To end up with the quasi-diagonalization, we have to add the expression
\begin{equation}
X\varepsilon \left( \mathbf{P,R}\right) +\varepsilon \left( \mathbf{P,R}%
\right) X^{+}
\end{equation}
Given the expression obtained previously for $X$, we have thus
\begin{eqnarray}
&&X\varepsilon \left( \mathbf{P,R}\right) +\varepsilon \left( \mathbf{P,R}%
\right) X^{+}  \notag \\
&=&-\frac{i}{4\hbar }\left[ \mathcal{A}_{R_{l}},\mathcal{A}_{P^{l}}\right]
\varepsilon \left( \mathbf{P,R}\right) -\frac{i}{4\hbar }\varepsilon \left(
\mathbf{P,R}\right) \left[ \mathcal{A}_{R_{l}},\mathcal{A}_{P^{l}}\right]
\end{eqnarray}
We can thus ultimately write the diagonalization process as :
\begin{eqnarray}
\left[ \left( U\left( \mathbf{P},\mathbf{R}\right) +X\right) H_{0}\left(
\mathbf{P,R}\right) \left( U^{+}\left( \mathbf{P,R}\right) +X^{+}\right) %
\right] &=&\varepsilon \left( \mathbf{P,R}\right) +\frac{1}{2}\left[
\mathcal{A}_{R_{l}}\nabla _{R^{l}}\varepsilon \left( \mathbf{P,R}\right)
+\nabla _{R^{l}}\varepsilon \left( \mathbf{P,R}\right) \mathcal{A}_{R_{l}}%
\right]  \notag \\
&&+\frac{1}{2}\left[ \mathcal{A}_{P^{l}}\nabla _{P_{l}}\varepsilon \left(
\mathbf{P,R}\right) +\nabla _{P_{l}}\varepsilon \left( \mathbf{P,R}\right)
\mathcal{A}_{P^{l}}\right]  \notag \\
&&-\frac{i}{2\hbar }\left[ \varepsilon \left( \mathbf{P,R}\right) ,\mathcal{A%
}_{P^{l}}\right] \mathcal{A}_{R_{l}}+\frac{i}{2\hbar }\left[ \varepsilon
\left( \mathbf{P,R}\right) ,\mathcal{A}_{R_{l}}\right] \mathcal{A}_{P^{l}}
\notag \\
&&+\frac{i}{4\hbar }\left[ \left[ \mathcal{A}_{R_{l}},\mathcal{A}_{P^{l}}%
\right] ,\varepsilon \left( \mathbf{P,R}\right) \right]  \label{DIAGH}
\end{eqnarray}
Let us conclude this section by noting that our transformation is not a
diagonalization at order $\hbar $, since it includes non diagonal
contributions of order $\hbar $ through the Berry connections $\mathcal{A}%
_{R_{l}}$ and $\mathcal{A}_{P^{l}}$, justifying the name
quasi-diagonalization. However, the next paragraph will show that these non
diagonal terms are only an artifact. Actually, projecting our transformed
Hamiltonian on the diagonal will in fact yield the genuine diagonalization.

\subsection{The ''Exact'' semiclassical diagonalization}

As mentioned in subsection B, our choice of transformation $U\left( \mathbf{%
\ P},\mathbf{R}\right) +XU$ is somewhat arbitrary, however it is sufficient
to perform the exact diagonalization as shown in the present paragraph.
Actually, since $U\left( \mathbf{P},\mathbf{R}\right) $ diagonalizes $H_{0}$
at zeroth order in $\hbar $, one can consider an-unknown- true
diagonalization unitary operator $U_{1}\left( \mathbf{P},\mathbf{R}\right) $
reducing to $U\left( \mathbf{P},\mathbf{R}\right) $ at zeroth order in $%
\hbar $. That is $U_{1}\left( \mathbf{P},\mathbf{R}\right) =U\left( \mathbf{P%
},\mathbf{R}\right) +\hbar \left( ...\right) $. As a consequence $%
U_{1}\left( \mathbf{P},\mathbf{R}\right) $ and $U\left( \mathbf{P},\mathbf{R}%
\right) +XU$ are equal at the zeroth order in $\hbar $ and the difference
\begin{equation}
\delta XU\equiv U_{1}\left( \mathbf{P},\mathbf{R}\right) -U\left( \mathbf{P},%
\mathbf{R}\right) -XU
\end{equation}
is of order $\hbar $. Moreover $U_{1}\left( \mathbf{P},\mathbf{R}\right) $
and $U\left( \mathbf{P},\mathbf{R}\right) +XU$ being both unitary, $\delta X$
is easily seen to be antihermitian. As a direct consequence, one can check
that the difference between the exact diagonalization and our approximate
one
\begin{equation}
\left[ \left( U\left( \mathbf{P},\mathbf{R}\right) +XU\right) H_{0}\left(
\mathbf{P,R}\right) \left( U^{+}\left( \mathbf{P,R}\right)
+U^{+}X^{+}\right) \right] -\left[ U_{1}\left( \mathbf{P},\mathbf{R}\right)
H_{0}\left( \mathbf{\ P,R}\right) U_{1}^{+}\left( \mathbf{P},\mathbf{R}%
\right) \right]
\end{equation}
is equal to
\begin{equation}
\left[ \delta X,\varepsilon \left( \mathbf{P,R}\right) \right]
\end{equation}
Given that $\varepsilon \left( \mathbf{P,R}\right) $ is diagonal, this last
term is always non diagonal. As a consequence, if we project on the diagonal
the difference between our transformations, one gets $0$.
\begin{equation}
\emph{P}\left[ \left( U\left( \mathbf{P},\mathbf{R}\right) +XU\right)
H_{0}\left( \mathbf{P,R}\right) \left( U^{+}\left( \mathbf{P,R}\right)
+U^{+}X^{+}\right) \right] -\emph{P}\left[ U_{1}\left( \mathbf{P},\mathbf{R}%
\right) H_{0}\left( \mathbf{P,R}\right) U_{1}^{+}\left( \mathbf{P},\mathbf{R}%
\right) \right] =0
\end{equation}
Here we have denoted $\emph{P}[...]$\ the projection on the diagonal.

Now, given that $U_{1}\left( \mathbf{P},\mathbf{R}\right) H_{0}\left(
\mathbf{P,R}\right) U_{1}^{+}\left( \mathbf{P},\mathbf{R}\right) $ is truly
diagonal, one has
\begin{equation}
\emph{P}\left[ U_{1}\left( \mathbf{P},\mathbf{R}\right) H_{0}\left( \mathbf{%
P,R}\right) U_{1}^{+}\left( \mathbf{P},\mathbf{R}\right) \right]
=U_{1}\left( \mathbf{P},\mathbf{R}\right) H_{0}\left( \mathbf{P,R}\right)
U_{1}^{+}\left( \mathbf{P},\mathbf{R}\right)
\end{equation}
so that ultimately
\begin{equation}
\emph{P}\left[ \left( U\left( \mathbf{P},\mathbf{R}\right) +XU\right)
H_{0}\left( \mathbf{P,R}\right) \left( U^{+}\left( \mathbf{P,R}\right)
+U^{+}X^{+}\right) \right] =U_{1}\left( \mathbf{P},\mathbf{R}\right)
H_{0}\left( \mathbf{P,R}\right) U_{1}^{+}\left( \mathbf{P},\mathbf{R}\right)
\end{equation}
We can therefore conclude, that the projection of our quasi-diagonalized
Hamiltonian, by eliminating thus the non diagonal parts, is in fact the
genuine diagonalized Hamiltonian at order $\hbar $.

\subsection{The Diagonal Hamiltonian}

From the previous discussion we understand that the genuine semiclassical
diagonal Hamiltonian $H_{D}$ is simply given by the projection on the
diagonal of the Hamiltonian Eq.\ref{DIAGH}:
\begin{eqnarray}
H_{D} &=&\emph{P}\left[ \left( U\left( \mathbf{P},\mathbf{R}\right)
+XU\right) H_{0}\left( \mathbf{P,R}\right) \left( U^{+}\left( \mathbf{P,R}%
\right) +U^{+}X^{+}\right) \right]   \notag \\
&=&\varepsilon \left( \mathbf{P,R}\right) +\frac{1}{2}\left[ \emph{A}%
_{R_{l}}\nabla _{R_{l}}\varepsilon \left( \mathbf{P,R}\right) +\nabla
_{R_{l}}\varepsilon \left( \mathbf{P,R}\right) \emph{A}_{R_{l}}\right] +%
\frac{1}{2}\left[ \emph{A}_{P_{l}}\nabla _{P_{l}}\varepsilon \left( \mathbf{%
P,R}\right) +\nabla _{P_{l}}\varepsilon \left( \mathbf{P,R}\right) \emph{A}%
_{P_{l}}\right]   \notag \\
&&+\emph{P}\left[ -\frac{i}{2\hbar }\left[ \varepsilon \left( \mathbf{P,R}%
\right) ,\mathcal{A}_{P_{l}}\right] \mathcal{A}_{R_{l}}+\frac{i}{2\hbar }%
\left[ \varepsilon \left( \mathbf{P,R}\right) ,\mathcal{A}_{R_{l}}\right]
\mathcal{\ A}_{P_{l}}\right]
\end{eqnarray}
where we introduced the notation $\emph{A}=\emph{P}\left[ \mathcal{A}\right]
.$ This Hamiltonian can be rewritten
\begin{eqnarray}
H_{D} &=&\varepsilon \left( \mathbf{P+}\emph{A}_{P},\mathbf{R+}\emph{A}%
_{R}\right) +\frac{i}{2\hbar }\emph{P}\left[ \left[ \varepsilon \left(
\mathbf{\ P,R}\right) ,\mathcal{A}_{R_{l}}\right] \mathcal{A}_{P_{l}}-\left[
\varepsilon \left( \mathbf{P,R}\right) ,\mathcal{A}_{P_{l}}\right] \mathcal{%
\ A}_{R_{l}}\right]   \notag \\
&\simeq &\varepsilon \left( \mathbf{p,r}\right) +\frac{i}{2\hbar }\emph{P}%
\left[ \left[ \varepsilon \left( \mathbf{p,r}\right) ,\mathcal{A}_{R_{l}}%
\right] \mathcal{A}_{P_{l}}-\left[ \varepsilon \left( \mathbf{p,r}\right) ,%
\mathcal{A}_{P_{l}}\right] \mathcal{A}_{R_{l}}\right]   \label{onebandH}
\end{eqnarray}
where we have defined the projected dynamical operators
\begin{eqnarray}
\mathbf{r} &=&\emph{P}\left[ \left( U\left( \mathbf{P},\mathbf{R}\right)
\right) \mathbf{R}U^{+}\mathbf{\left( \mathbf{P,R}\right) }\right] =\mathbf{R%
}+\emph{A}_{\mathbf{R}}  \notag \\
\mathbf{p} &=&\emph{P}\left[ U\left( \mathbf{P},\mathbf{R}\right) \mathbf{P}%
U^{+}\mathbf{\left( \mathbf{P,R}\right) }\right] =\mathbf{P}+\emph{A}_{%
\mathbf{P}}
\end{eqnarray}
The non-canonical dynamical variables $\left( \mathbf{p,r}\right) $ have
corrections of order $\hbar $ through the presence of the Berry connections.

\subsection{The equations of motion}

Given the Hamiltonian derived in the previous subsection, the equations of
motion can now be easily derived. As usual \cite{ALAIN,PIERRE} the dynamics
has to be considered, not for the usual position $\mathbf{R}$ and momentum $%
\mathbf{P}$, but rather for the projected variables $\mathbf{r}$ and $%
\mathbf{p}$. These new dynamical operators which naturally appear in our
diagonalization process at the $\hbar $ order have components which do not
commute any more. Actually
\begin{eqnarray}
\left[ r_{i},r_{j}\right]  &=&i\Theta _{ij}^{rr}=i\hbar \left( \nabla
_{P_{i}}\emph{A}_{R_{j}}-\nabla _{P_{j}}\emph{A}_{R_{i}}\right) +\left[
\emph{A}_{R_{j}},\emph{A}_{R_{i}}\right]   \notag \\
\left[ p_{i},p_{j}\right]  &=&i\Theta _{ij}^{pp}=-i\hbar \left( \nabla
_{R_{i}}\emph{A}_{P_{j}}-\nabla _{R_{j}}\emph{A}_{P_{i}}\right) +\left[
\emph{A}_{P_{i}},\emph{A}_{P_{j}}\right]   \notag \\
\left[ p_{i},r_{j}\right]  &=&-i\hbar \delta _{ij}+i\Theta
_{ij}^{pr}=-i\hbar \delta _{ij}-i\hbar \left( \nabla _{R_{i}}\emph{A}%
_{R_{j}}+\nabla _{P_{j}}\emph{A}_{P_{i}}\right) +\left[ \emph{A}_{P_{i}},%
\emph{A}_{R_{j}}\right]   \label{commutationrelations}
\end{eqnarray}
the $\Theta _{ij}$ being the so called non-Abelian Berry curvatures. They
are of order $\hbar ^{2}$ as lowest order corrections to the commutators,
but will actually induce semiclassical corrections of order $\hbar $ to the
equations of motion. Indeed the one band-Hamiltonian Eq. \ref{onebandH}
yields directly to general equations of motion for $\mathbf{r}$, $\mathbf{p}$
:
\begin{eqnarray}
\mathbf{\dot{r}} &=&\frac{i}{\hbar }\left[ \mathbf{r},\varepsilon \left(
\mathbf{p,r}\right) \right] +\frac{i}{\hbar }\left[ \mathbf{r},\frac{i}{%
2\hbar }\emph{P}\left[ \left[ \varepsilon \left( \mathbf{p,r}\right) ,%
\mathcal{A}_{R_{l}}\right] \mathcal{A}_{P^{l}}-\left[ \varepsilon \left(
\mathbf{p,r}\right) ,\mathcal{A}_{P^{l}}\right] \mathcal{A}_{R_{l}}\right] %
\right]   \notag \\
\mathbf{\dot{p}} &=&\frac{i}{\hbar }\left[ \mathbf{p,}\varepsilon \left(
\mathbf{p,r}\right) \right] +\frac{i}{\hbar }\left[ \mathbf{p},\frac{i}{%
2\hbar }\emph{P}\left[ \left[ \varepsilon \left( \mathbf{p,r}\right) ,%
\mathcal{A}_{R_{l}}\right] \mathcal{A}_{P^{l}}-\left[ \varepsilon \left(
\mathbf{p}^{+}\mathbf{,r}^{+}\right) ,\mathcal{A}_{P^{l}}\right] \mathcal{A}%
_{R_{l}}\right] \right]
\end{eqnarray}
where the commutators can be computed through the previous commutation rules
Eq.\ref{commutationrelations}. The last term in each equation represents a
contribution of ''magnetization'' type (see the following applications) and
has the advantage to present this general form whatever the system initially
considered. However, to put some flesh on these equations, we now turn to
several examples covered by our formalism.

\section{Application 1: The Dirac electron.}

To apply our previous formalism, we will consider two cases of Dirac
Hamiltonians : The electromagnetic field and the static symmetrical
gravitational field. These two cases have already been treated by different
methods (\cite{BLIOKH1} and \cite{OBUKHOV,SILENKO}), but in the second case
(gravitational field) references to Berry phases was made for the first time
in\textbf{\ }\cite{PHOTONPIERRE}\textbf{. }

\subsection{The Dirac electron in an electromagnetic field}

The diagonalization of the Dirac Hamiltonian in the presence of an
electromagnetic field is a difficult problem which was solved only
approximately in the non-relativistic limit in an $m^{-1}$ expansion.
Another approach consists in diagonalizing the Hamiltonian at the
semi-classical order as was done in \cite{BLIOKH1} using an approximate
Foldy-Wouthuysen transformation \cite{FW}. From the semi-classical
Hamiltonian the equations of motion were derived showing a topological
spin-transport effect due to the presence of the Berry phases. Here we
propose to apply our general formalism for the semi-classical
diagonalization of the Dirac Hamiltonian to show the effectiveness of our
general method. It is worth noticing that the method developed above has now
to be adapted as we will transform the Dirac Hamiltonian into a ($2\times 2)$%
\ block-diagonal matrix (due to the spin degree of freedom). However, it is
easy to check that, since the non diagonal components for the energy blocks
are of order $\hbar ,$\ these corrections do not impair our general formulas
for the Berry phases contributions\textbf{.}

We thus start with the following Dirac Hamiltonian:
\begin{equation}
H_{0}\left( \mathbf{P},\mathbf{R}\right) =\mathbf{\alpha .}\left( \mathbf{P-A%
}(\mathbf{R})\right) +\beta m+V(\mathbf{R})
\end{equation}
where the matrix $\mathbf{\alpha }$ and $\beta $ are the usual \textbf{(}$%
4\times 4)$ Dirac matrices:
\begin{equation}
\alpha _{i}=\left(
\begin{array}{cc}
\begin{array}{cc}
0 & 0 \\
0 & 0
\end{array}
& \sigma _{i} \\
\sigma _{i} &
\begin{array}{cc}
0 & 0 \\
0 & 0
\end{array}
\end{array}
\right)
\end{equation}
with $\sigma _{i}$ the usual \textbf{(}$2\times 2)$ Pauli matrices ($i=1,2,3$%
) and
\begin{equation}
\beta =\left(
\begin{array}{cccc}
1 & 0 & 0 & 0 \\
0 & 1 & 0 & 0 \\
0 & 0 & -1 & 0 \\
0 & 0 & 0 & -1
\end{array}
\right)
\end{equation}
Replacing $\mathbf{R}$ by the parameter $\widetilde{\mathbf{r}}$, $\mathbf{A}%
(\widetilde{\mathbf{r}})$ just shifts the momentum, so that we can
diagonalize $H_{0}\left( \mathbf{P},\widetilde{\mathbf{r}}\right) $ through
the well known Foldy-Wouthuysen transformation
\begin{equation}
U\left( \mathbf{P,}\widetilde{\mathbf{r}}\right) =\frac{E+m+\beta \mathbf{%
\alpha .}\left( \mathbf{P-A}(\widetilde{\mathbf{r}})\right) }{\sqrt{2E(E+m)}}
\end{equation}
where$E=\sqrt{\left( \mathbf{P-A}(\widetilde{\mathbf{r}})\right) ^{2}+m^{2}}$%
. In this context, introducing the dependence in $\mathbf{R}$ we define the
(non projected) Berry connections at first order in $\hbar $ as a ($4\times
4)$\ matrix
\begin{equation}
\mathcal{A}_{R_{i}}=i\hbar U\nabla _{P_{i}}U^{+}=\hbar \frac{i\mathbf{\alpha
.}\left( \mathbf{P-A}(\mathbf{R)}\right) P_{i}\beta +i\beta E(E+m)\alpha
_{i}-E\left( \mathbf{\Sigma }\times \left( \mathbf{P-A}(\mathbf{R)}\right)
\right) _{i}}{2E^{2}(E+m)}
\end{equation}
and
\begin{equation}
\mathcal{A}_{P_{l}}=-i\hbar U\nabla _{R_{l}}U^{+}=\nabla _{R_{l}}A_{k}%
\mathbf{(R)}\mathcal{A}_{R_{k}}
\end{equation}
where $E$ will now denote the symmetrized form of $E=\sqrt{\left( \mathbf{P-}%
A(\mathbf{R)}\right) ^{2}+m^{2}}$ (here at zeroth order in $\hbar $ we can
consider that $\mathbf{P}$ and $\mathbf{R}$ commute). The spin matrices $%
\Sigma _{i}$ ($i=1,2,3$) are given by\textbf{\ }
\begin{equation*}
\mathbf{\Sigma }_{i}\mathbf{=}\left(
\begin{array}{cc}
\sigma _{i} &
\begin{array}{cc}
0 & 0 \\
0 & 0
\end{array}
\\
\begin{array}{cc}
0 & 0 \\
0 & 0
\end{array}
& \sigma _{i}
\end{array}
\right)
\end{equation*}
The general method developed in the previous section allows us to write the
diagonal Hamiltonian $H_{D}$\ as a matrix generalization of Eq.\ref{onebandH}
\begin{equation}
H_{D}=\emph{P}\left[ UH_{0}U^{+}\right] =\beta \varepsilon \left( \mathbf{p,r%
}\right) +\frac{i}{2\hbar }\emph{P}\left[ \left[ \varepsilon \left( \mathbf{%
p,r}\right) ,\mathcal{A}_{R_{l}}\right] \mathcal{A}_{P^{l}}-\left[
\varepsilon \left( \mathbf{p,r}\right) ,\mathcal{A}_{P^{l}}\right] \mathcal{A%
}_{R_{l}}\right] +V(\mathbf{r})  \label{energydirac}
\end{equation}
with $\varepsilon \left( \mathbf{p,r}\right) =\sqrt{\left( \mathbf{p-A}(%
\mathbf{r)}\right) ^{2}+m^{2}}$. In Eq. \ref{energydirac} the operators $%
\left( \mathbf{p,r}\right) $ are the physical dynamical variables satisfying
the non-canonical commutations relations Eq. \ref{commutationrelations}.
Using the expressions for the Berry connections, $H_{D}$ can be rewritten
as:
\begin{equation}
H_{D}=\beta \varepsilon \left( \mathbf{p,r}\right) +\emph{P}\left[ -\frac{i}{%
2}\left[ \varepsilon \left( \mathbf{p,r}\right) ,U\nabla _{P^{i}}U^{+}\right]
\varepsilon ^{ijk}U\nabla _{P^{j}}U^{+}\right] \frac{B^{k}(\mathbf{\mathbf{r}%
)}}{\hbar }+V(\mathbf{r})
\end{equation}
Moreover, a straightforward computation shows that one can write $\emph{P}%
\left[ -\frac{i}{2}\left[ \varepsilon \left( \mathbf{p,r}\right) ,U\nabla
_{P^{i}}U^{+}\right] .\varepsilon ^{ijk}U\nabla _{P_{j}}U^{+}\right] \frac{%
B^{k}(\mathbf{\mathbf{r})}}{\hbar }=\beta \frac{\hbar \mathbf{\Sigma }.%
\mathbf{B}}{2E}-\beta \frac{\mathbf{L}.\mathbf{B}}{E}$ at the first order in
$\hbar ,$ where we have introduced the intrinsic angular momentum of
semiclassical particles $\mathbf{L=P\times }\emph{A}_{\mathbf{R}}$ with $%
\emph{A}_{\mathbf{R}}=\emph{P}\left[ \mathcal{A}_{R}\right] =\hbar \frac{%
\left( \mathbf{P-A}(\mathbf{R)}\right) \times \mathbf{\Sigma }}{2E(E+m)}$
the projection of the Berry connection on the diagonal\textbf{. }As a
consequence, the Hamiltonian to be considered is given by
\begin{equation}
H_{D}=\beta \varepsilon \left( \mathbf{p,r}\right) +\beta \frac{\hbar
\mathbf{\Sigma }.\mathbf{B}}{2E}-\beta \frac{\mathbf{L}.\mathbf{B}}{E}+V(%
\mathbf{r})
\end{equation}
which is the Hamiltonian deduced in \cite{BLIOKH1} from a different approach
and which leads of course to the dynamics described in that paper.

\subsection{The electron in a static gravitational field}

The behavior of Dirac particles in static gravitational field is an
important issue, at the crossroad of particle physics and cosmology.
Different approaches for the diagonalization of the Hamiltonian leads to
contradictory results in particular with regard to the existence of a dipole
spin-gravity coupling \cite{OBUKHOV, SILENKO}. It is not our goal to discuss
this specific point but we study the semiclassical diagonalization of the
Hamiltonian to get the velocity and momentum evolution. We can in particular
compare our results with the article \cite{SILENKO} who uses a
Foldy-Wouthuysen transformation instead of a semiclassical approximation.

The interaction of a Dirac particle with a symmetric static gravitational
field ($g_{00}=V(\mathbf{R}),g_{i0}=0,g_{ij}=\delta _{ij}F(\mathbf{R})$) is
described by the Hamiltonian \cite{SILENKO}
\begin{equation}
H_{0}=\frac{1}{2}\left( \mathbf{\alpha }.\mathbf{P}F(\mathbf{R})+F(\mathbf{R}%
)\mathbf{\alpha }.\mathbf{P}\right) +\beta mV(\mathbf{r})
\end{equation}
The Foldy Whouthuysen transformation when $\mathbf{R}$ is replaced by a
parameter $\widetilde{\mathbf{r}}$ is given by
\begin{equation}
U\left( \mathbf{P,}\widetilde{\mathbf{r}}\right) =\frac{E+mV(\widetilde{%
\mathbf{r}})+\beta F(\widetilde{\mathbf{r}})\mathbf{\alpha .P}}{\sqrt{%
2E(E+mV(\widetilde{\mathbf{r}}))}}
\end{equation}
with $E\left( \mathbf{P,}\widetilde{\mathbf{r}}\right) =\sqrt{F^{2}(%
\widetilde{\mathbf{r}})\mathbf{P}^{2}+m^{2}V^{2}(\widetilde{\mathbf{r}})}$.
This is quite the same as the free particle transformation. As a consequence
introducing again the $\mathbf{R}$ dependence yields the non projected Berry
connections for the position and the momentum operators :
\begin{eqnarray}
\mathcal{A}_{R_{i}} &=&i\hbar U\nabla _{P_{i}}U^{+}=\hbar \frac{iF^{3}(%
\mathbf{R})\mathbf{\alpha .P}P_{i}\beta +i\beta F(\mathbf{R})\varepsilon
(\varepsilon +mV(\mathbf{R}))\alpha _{i}-\varepsilon F^{2}(\mathbf{R})\left(
\mathbf{\Sigma }\times \mathbf{P}\right) _{i}}{2\varepsilon ^{2}\left(
\varepsilon +mV(\mathbf{R})\right) }  \notag \\
\mathcal{A}_{P_{i}} &=&-i\hbar U\nabla _{P_{i}}U^{+}=-\hbar i\frac{m\left(
\mathbf{\nabla }_{R_{i}}\phi \right) \beta F^{2}(\mathbf{R})\mathbf{\alpha .P%
}}{2\varepsilon ^{2}}
\end{eqnarray}
with $\phi =\frac{V}{F}$, $\mathbf{\Sigma }$ the spin of the electron\ and
the energy $\varepsilon $\ is now given by $\varepsilon \left( \mathbf{P,R}%
\right) =\sqrt{\left( F^{2}(\mathbf{R})\mathbf{P}^{2}+\mathbf{P}^{2}F^{2}(%
\mathbf{R})\right) /2+m^{2}V^{2}(\mathbf{R})}.$

The expressions for $\mathcal{A}_{R_{i}}$ and $\mathcal{A}_{P_{i}}$ allow us
ultimately to define the semiclassical transformation : $U\left( \mathbf{P,R}%
\right) +\frac{i}{4\hbar }\left[ \mathcal{A}_{P^{l}},\mathcal{A}_{R_{l}}%
\right] U\left( \mathbf{P,R}\right) $ and to compute the diagonal
Hamiltonian
\begin{equation}
H_{D}=\emph{P}\left[ UH_{0}U^{+}\right] =\beta \varepsilon \left( \mathbf{p,r%
}\right) -\beta \frac{F^{3}(\mathbf{r})}{2\varepsilon ^{2}}m\hbar \mathbf{%
\nabla }\phi (\mathbf{r}).\left( \mathbf{p}\times \mathbf{\Sigma }\right)
\label{Hplus}
\end{equation}
where the dynamical variables which are deduced from the projections $\emph{A%
}_{\mathbf{R}}=\emph{P}\left[ \mathcal{A}_{\mathbf{R}}\right] =\hbar \frac{%
F^{2}\mathbf{\Sigma }\times \mathbf{P}}{2\varepsilon (\varepsilon +mV)}$ and
$\emph{A}_{\mathbf{P}}=\emph{P}\left[ \mathcal{A}_{\mathbf{P}}\right] =0$
are
\begin{eqnarray}
\mathbf{r} &=&\mathbf{R+}\emph{A}_{\mathbf{R}}=\mathbf{R-}\hbar \frac{F^{2}(%
\mathbf{R})\mathbf{\Sigma }\times \mathbf{P}}{2\varepsilon (\varepsilon +mV(%
\mathbf{R}))}  \notag \\
\mathbf{p} &=&\mathbf{P+}\emph{A}_{\mathbf{P}}=\mathbf{P}
\end{eqnarray}
The commutators between these variables are thus
\begin{eqnarray}
\left[ r_{i},r_{j}\right] &=&i\Theta _{ij}^{rr}  \notag \\
\left[ p_{i},r_{j}\right] &=&-i\hbar \delta _{ij}+i\Theta _{ij}^{pr}  \notag
\\
\left[ p_{i},p_{j}\right] &=&0
\end{eqnarray}
with the Berry curvatures
\begin{eqnarray}
\Theta _{ij}^{rr} &=&-\frac{\hbar ^{2}F^{3}(\mathbf{r})\varepsilon ^{ijk}}{%
2\varepsilon ^{3}\left( \mathbf{p,r}\right) }\left( m\phi (\mathbf{r})%
\mathbf{\Sigma }_{k}+\frac{F(\mathbf{r})\left( \mathbf{\Sigma .p}\right)
\mathbf{p}_{k}}{\varepsilon \left( \mathbf{p,r}\right) +mV(\mathbf{r})}%
\right)  \notag \\
\Theta _{ij}^{pr} &=&-\frac{\hbar ^{2}F^{3}(\mathbf{r})}{2\varepsilon
^{3}\left( \mathbf{p,r}\right) }m\nabla _{i}\phi (\mathbf{r})\left( \mathbf{%
\Sigma }\times \mathbf{p}\right) _{j}  \notag \\
\Theta _{ij}^{pp} &=&0  \label{energy}
\end{eqnarray}
One can check, after developing $\mathbf{r}$ as a function of $\mathbf{R}$
and the Berry connection, that the Hamiltonian Eq. \ref{Hplus} coincides
with the one given in \cite{SILENKO} at order $\hbar $. This confirms also
the validity of the Foldy Wouthuysen approach asserted in \cite{SILENKO} in
opposition with the transformation proposed in \cite{OBUKHOV}. However our
approach is more general since it does not require an expansion in $V$ and $%
F $ as done in \cite{SILENKO}. Of course, we retrieve the result of \cite
{SILENKO} if we expand expression \ref{energy} at the leading order in $F$
and $V$. Note also that when $m=0$ one recovers the Hamiltonian for the
Neutrino or the photon proposed in \cite{ALAIN}\cite{PHOTONPIERRE}.

To conclude this paragraph, we can derive the equations of motion with the
help the noncanonical commutators between the coordinates and the spin
\begin{eqnarray}
\Theta _{ij}^{r\Sigma }=\left[ r_{i},\Sigma _{j}\right] &=&i\hbar c^{2}\frac{%
-p_{j}\mathbf{\Sigma }_{i}+\mathbf{p.\Sigma }\delta _{ij}}{\varepsilon
\left( \mathbf{p,r}\right) \left( \varepsilon \left( \mathbf{p,r}\right) +mV(%
\mathbf{r})\right) }  \notag \\
\Theta _{ij}^{p\Sigma }=\left[ p_{i},\Sigma _{j}\right] &=&0
\end{eqnarray}
We then can deduce the semiclassical equations of motion for the electron in
a symmetric static gravitational field (by projecting on the positive energy
subspace)
\begin{eqnarray}
\mathbf{\dot{r}} &=&\left( 1-\frac{\Theta ^{pr}}{\hbar }\right) \nabla _{%
\mathbf{p}}\varepsilon -\frac{1}{\hbar }\mathbf{\dot{p}\times }\Theta ^{rr}+%
\frac{i}{\hbar }\nabla _{\mathbf{\Sigma }}\varepsilon .\Theta ^{r\Sigma }
\notag \\
\mathbf{\dot{p}} &=&-\left( 1-\frac{\Theta ^{pr}}{\hbar }\right) \nabla _{%
\mathbf{r}}\varepsilon  \label{velocity}
\end{eqnarray}
where we have defined the vectors $\Theta ^{rr}$ and $\Theta ^{rp}$ through $%
\Theta _{ij}^{rr}=\varepsilon ^{ijk}\Theta _{k}^{rr}$ and $\Theta
_{ij}^{pr}=\varepsilon ^{ijk}\Theta _{k}^{rp}.$

The velocity equation in Eq.\ref{velocity} contains in particular an
anomalous velocity term $\mathbf{\dot{p}\times }\Theta ^{rr}/\hbar $ of
order $\hbar $ which causes an additional displacement of the electrons
orthogonally to the momentum $\mathbf{p}$. The dynamics of the system must
be completed by the spin dynamics which is

\begin{equation}
\hbar \mathbf{\dot{\Sigma}}=\frac{mV(\mathbf{r})\hbar }{\varepsilon
(\varepsilon +mV(\mathbf{r}))}\mathbf{\Sigma \times }\left( \mathbf{\nabla }%
V(\mathbf{r})\mathbf{\times P}\right) -\frac{\hbar }{\varepsilon }\mathbf{%
\Sigma \times }\left( \mathbf{\nabla }F(\mathbf{r})\mathbf{\times P}\right)
\end{equation}
Although the position dynamics differs from the one of obtained in \cite
{SILENKO}, due to our choice for the physical position operator ($\mathbf{r}$
instead of $\mathbf{R}$), one can show that the equations for $\dot{p}$ and $%
\hbar \dot{\Sigma}$\ reduce to \cite{SILENKO} in the case of weak fields.

\section{Application 2 : The electron in a periodic potential}

This application has already been independently studied in \cite{PIERRE},
and is easily recovered by the present general setup. The purpose is to find
the semiclassical Hamiltonian for an electron in a periodic potential facing
an electromagnetic field. This topic was also already dealt with in \cite
{NIU1} in the context of wave packets dynamics. We will show that the
semiclassical equations of motions which are very essential in the solid
state physics context must be corrected by Berry phases terms. To apply our
formalism, consider an electron in an crystal lattice perturbated by the
presence of an external electromagnetic field. As is usual, we express the
total magnetic field as the sum of a constant field $\mathbf{B}$ and small
nonuniform part $\delta \mathbf{B}(\mathbf{R})$. The Schr$\overset{..}{\text{%
o}}$dinger equation reads $\left( H_{0}-e\phi (\mathbf{R})\right) \Psi (%
\mathbf{R})=E\Psi (\mathbf{R})$ with $H_{0}$ the magnetic contribution ($%
\phi $ being the electric potential) which reads
\begin{equation}
H_{0}=\left( \frac{\mathbf{P}}{2m}+e\mathbf{A}(\mathbf{R})+e\delta \mathbf{A}%
(\mathbf{R})\right) ^{2}+V(\mathbf{R}),\text{ \ \ \ \ }\mathbf{P=-}i\hbar
\nabla  \label{Hmagnetic}
\end{equation}
where $\mathbf{A}(\mathbf{R})$ and $\delta \mathbf{A}(\mathbf{R})$ are the
vectors potential of the homogeneous and inhomogeneous magnetic field,
respectively, and $V(\mathbf{R})$ the periodic potential. The large constant
part $\mathbf{B}$ is chosen such that the magnetic flux through a unit cell
is a rational fraction of the flux quantum $h/e$. The advantage of such a
decomposition is that for $\delta \mathbf{A}(\mathbf{R})=0$ the magnetic
translation operators $\mathbf{T}(\mathbf{b})=\exp (i\mathbf{K}.\mathbf{b})$
defined in Appendix $2$, with $\mathbf{K}$ the generator of translation, are
commuting quantities allowing to exactly diagonalize the Hamiltonian and to
treat $\delta \mathbf{A}(\mathbf{R})$ as a small perturbation. The state
space of the Bloch electron is spanned by the basis vectors of plane waves%
\textbf{\ }$\left| n,\mathbf{k}\right\rangle =\left| \mathbf{k}\right\rangle
\otimes \left| n\right\rangle $ with $n$ corresponding to a band index. The
state $\left| n\right\rangle $ can be seen as a canonical base vector $%
\left| n\right\rangle =(0...010...0...)$ (with $1$ at the $n$th position)
such that $U^{+}\left( \mathbf{k}\right) \left| n\right\rangle =\left|
u_{n}\left( \mathbf{k}\right) \right\rangle $ with $\left| u_{n}\left(
\mathbf{k}\right) \right\rangle $ the periodic part of the magnetic Bloch
waves.\textbf{\ }In this representation $\mathbf{K}\left| n,\mathbf{k}%
\right\rangle =\mathbf{k}\left| n,\mathbf{k}\right\rangle $ and consequently%
\textbf{\ }the position operator is $\mathbf{R=}i\partial /\partial \mathbf{k%
}$, implying the canonical commutation relations $\left[ \mathbf{R}_{i}%
\mathbf{,K}_{j}\right] =i\delta _{ij}$.

We first perform the diagonalization of the Hamiltonian in Eq.\ref{Hmagnetic}
for $\delta \mathbf{A}=0$ by diagonalizing simultaneously $H_{0}$ and the
magnetic translation operators $\mathbf{T}$. The diagonalization is
performed as follows: start with an arbitrary basis of eigenvectors of $%
\mathbf{T}$. As explained in Appendix $2$, in this basis $H_{0}$ can be seen
as a square matrix with operators entries.$H_{0}$ is diagonalized through a
unitary matrix $U(\mathbf{K})$ which should depend only on $\mathbf{K}$
(since $U$ should leave $\mathbf{K}$ invariant, i.e., $U\mathbf{K}U^{+}=%
\mathbf{K}$) and whose precise expression is not necessary for the
derivation of the equations of motion, such that $UHU^{+}=\mathcal{E}(%
\mathbf{K})-e\phi (U\mathbf{R}U^{+})$, where $\mathcal{E}(\mathbf{K})$ is
the diagonal energy matrix made of elements $\mathcal{E}_{n}(\mathbf{K})$
with $n$ the band index (i.e. the diagonal representation of $H_{0}$).

Now, to add a perturbation $\delta A(\mathbf{R)}$ as in (\cite{PIERRE}),
that breaks the translational symmetry, we have to replace $\mathbf{K}$ in
all expressions by
\begin{equation}
\mathbf{\tilde{K}}=\mathbf{K}+e\frac{\delta A(\mathbf{R)}}{\hbar }
\end{equation}
and as the flux $\mathbf{\delta B}$ on a plaquette is not a rational
multiple of the flux quantum, we cannot diagonalize simultaneously its
components $\tilde{K}_{i}$ since they do not commute anymore. Actually
\begin{equation}
\hbar \lbrack \tilde{K}^{i},\tilde{K}^{j}]=-ie\varepsilon ^{ijk}\delta B_{k}(%
\mathbf{R})
\end{equation}
As a consequence of this non-commutativity, we just aim at
quasi-diagonalizing our Hamiltonian at the semiclassical order (with
accuracy $\hbar $). To do that we replace $U(\mathbf{K})$ by $U\left(
\mathbf{\tilde{K}}\right) $, so that the non projected Berry connections are
$\mathcal{A}_{R_{i}}=iU\nabla _{\widetilde{K}_{i}}U^{+}$ and $\mathcal{A}%
_{K_{l}}=\nabla _{R_{l}}\delta A_{k}\mathbf{(R)}\mathcal{A}_{R_{k}}.$ From
these we can define the $n$th intraband position and momentum operators $%
\mathbf{r}_{n}\mathbf{=R+}\emph{A}_{n}$ and $\mathbf{\tilde{k}}_{n}\simeq
\tilde{\mathbf{K}}-e\emph{A}_{n}(\tilde{\mathbf{k}}_{n})\times \delta
\mathbf{B}(\mathbf{r}_{n})/\hbar +O(\hbar )$ with $\emph{A}_{n}=\emph{P}%
_{n}(U\nabla _{\widetilde{\mathbf{K}}}U^{+})$ the projection of the Berry
connection on the chosen $n$th Band \cite{PIERRE}. It can be readily seen
that the matrix elements of\textbf{\ }$\emph{A}_{n}$\textbf{\ }can be
written $\emph{A}_{n}\left( \mathbf{k}\right) =i\left\langle u_{n}\left(
\mathbf{k}\right) \right| \nabla _{\mathbf{k}}\left| u_{n}\left( \mathbf{k}%
\right) \right\rangle $\textbf{\ (}see also ref. \cite{LANDAU}for the
derivation of the position operator in the diagonal representation). What is
totally new here is the transformation on the momentum operator $\tilde{k}%
_{n}$\ which get also a Berry connection correction.

Using our results of section 2, the full Hamiltonian Eq. \ref{Hmagnetic} can
thus be diagonalized through the transformation $U(\mathbf{\tilde{K}})+\frac{%
i}{4\hbar }\left[ \mathcal{A}_{R_{l}},\mathcal{A}_{P^{l}}\right] U(\mathbf{%
\tilde{K}})$ plus a projection on the chosen $n$-th Band as it is usual in
solid state physics (the so called one band approximation)
\begin{eqnarray}
\emph{P}_{n}\left[ U\left( \mathbf{\tilde{K}}\right) HU^{+}\left( \mathbf{%
\tilde{K}}\right) \right] &=&\emph{P}_{n}\left[ \mathcal{E}\left( \mathbf{%
\tilde{k}}\right) -\frac{i}{4}\left[ \mathcal{E}(\mathbf{K}),U\nabla
_{K_{i}}U^{+}\right] \varepsilon ^{ijk}\frac{\delta B^{k}(\mathbf{r)}}{\hbar
}U\nabla _{K_{j}}U^{+}\right.  \notag \\
&&\left. -\frac{i}{4}U\nabla _{K_{j}}U^{+}\left[ \mathcal{E}(\mathbf{K}%
),U\nabla _{K_{i}}U^{+}\right] \varepsilon ^{ijk}\frac{\delta B^{k}(\mathbf{%
r)}}{\hbar }\right]  \notag \\
&=&\mathcal{E}_{n}\left( \mathbf{\tilde{k}}_{n}\right) -\mathcal{M}(\tilde{%
\mathbf{K}}).\delta \mathbf{B}(\mathbf{\mathbf{r}}_{n}\mathbf{)+}O\mathbf{(}%
\hbar ^{2}\mathbf{)}  \label{ENER}
\end{eqnarray}
where the energy levels $\mathcal{E}_{n}\left( \mathbf{\tilde{k}}_{n}\right)
$ are the same as $\mathcal{E}_{n}(\mathbf{K})$ with $\mathbf{\tilde{k}}_{n}$
replacing $\mathbf{K.}$ The magnetization $\mathcal{M}(\tilde{\mathbf{K}})=%
\emph{P}_{n}(\frac{ie}{2\hbar }\left[ \mathcal{E}(\tilde{\mathbf{K}}),%
\mathcal{A}(\tilde{\mathbf{K}})\right] \times \mathcal{A}(\tilde{\mathbf{K}}%
))$ can be written under the usual form \cite{LANDAU} in the $(\mathbf{k},n)$
representation
\begin{equation*}
\mathcal{M}_{nn}^{i}=\frac{ie}{2\hbar }\varepsilon ^{ijk}\sum_{n^{\prime
}\neq n}(\mathcal{E}_{n}-\mathcal{E}_{n^{\prime }})(\mathcal{A}%
_{j})_{nn^{\prime }}(\mathcal{A}_{k})_{n^{\prime }n}
\end{equation*}
We mention that this magnetization (the orbital magnetic moment of Bloch
electrons), has been obtained previously in the context of electron wave
packets dynamics \cite{NIU1}.

From the expression of the energy Eq. \ref{ENER} we can deduce the equations
of motion (with the band index $n$ now omitted)
\begin{eqnarray}
\dot{\mathbf{r}} &=&\partial E(\tilde{\mathbf{k}})/\hbar \partial \tilde{%
\mathbf{k}}-\dot{\tilde{\mathbf{k}}}\times \Theta (\tilde{\mathbf{k}})
\notag \\
\hbar \dot{\tilde{\mathbf{k}}} &=&-e\mathbf{E}-e\dot{\mathbf{r}}\times
\delta \mathbf{B}(\mathbf{r})-\mathcal{\mathbf{M}}\partial \delta \mathbf{B}%
/\partial \mathbf{r}  \label{EQM}
\end{eqnarray}
where $\left[ r^{i},r^{j}\right] =i\Theta ^{ij}(\widetilde{\mathbf{k}})$
with $\Theta ^{ij}(\widetilde{\mathbf{k}})=\partial ^{i}\mathcal{A}^{j}(%
\widetilde{\mathbf{k}})-\partial ^{j}\mathcal{A}^{i}(\widetilde{\mathbf{k}})$
the Berry curvature.

As explained in \cite{PIERRE} these equations are the same as the one
derived in \cite{NIU1} from a completely different formalism.

\section{Conclusion}

Some recent applications of semi classical methods to several branches of
Physics, such as spintronics or solid state physics have shown the relevance
of Berry Phases contributions to the dynamics of a system. However, these
progresses called for a rigorous Hamiltonian treatment that would allow for
deriving naturally the role of the Berry phase.

This paper has been devoted to derive a semi classical diagonalization
method for a broad class of quantum systems, including the electron in a
periodic potential and the Dirac Hamiltonian. Doing so, we have exhibited a
general pattern for this class of systems implying the role of the Berry
phases both for the position and the momentum. In such a context, the
coordinates and momenta algebra are no longer commutative, and the dynamical
equations for these variables directly include the influence of Berry phases
through the parameters of noncommutativity (Berry curvatures) and through an
abstract magnetization term. Applications of our formalism consider the
Dirac electron in an electromagnetic field, or in a particular case of
static gravitational field, as well as the electron in a periodic potential.
Our results are promising and indicate that our method will probably apply
to several other systems.

\bigskip

\textbf{Acknowledgment.} The authors wish to thank Aileen Lotz for a
critical reading of the manuscript.

\subsection{Appendix 1}

\begin{quote}
To start our computation, we will use the formula given in the preliminary :
\begin{eqnarray}
\frac{i}{2}\hbar Asym\nabla _{R_{l}}\nabla _{P^{l}}\left[ U\left( \mathbf{P},%
\mathbf{R}\right) H_{0}\left( \mathbf{P,R}\right) U^{+}\left( \mathbf{P,R}%
\right) \right] &=&\frac{i}{2}\hbar Asym\nabla _{R_{l}}\nabla _{P^{l}}\left[
U\left( \mathbf{P},\mathbf{R}\right) \right] H_{0}\left( \mathbf{P,R}\right)
U^{+}\left( \mathbf{P,R}\right)  \notag \\
&&+\frac{i}{2}\hbar U\left( \mathbf{P},\mathbf{R}\right) Asym\nabla
_{R_{l}}\nabla _{P^{l}}\left[ H_{0}\left( \mathbf{P,R}\right) \right]
U^{+}\left( \mathbf{P,R}\right)  \notag \\
&&+\frac{i}{2}\hbar U\left( \mathbf{P},\mathbf{R}\right) H_{0}\left( \mathbf{%
P,R}\right) Asym\nabla _{R_{l}}\nabla _{P^{l}}\left[ U^{+}\left( \mathbf{P,R}%
\right) \right]  \notag \\
&&+\frac{i}{2}\hbar \nabla _{R_{l}}U\left( \mathbf{P},\mathbf{R}\right)
\nabla _{P^{l}}H_{0}\left( \mathbf{P,R}\right) U^{+}\left( \mathbf{P,R}%
\right)  \notag \\
&&-\frac{i}{2}\hbar \nabla _{P^{l}}U\left( \mathbf{P},\mathbf{R}\right)
\nabla _{R_{l}}H_{0}\left( \mathbf{P,R}\right) U^{+}\left( \mathbf{P,R}%
\right)  \notag \\
&&-\frac{i}{2}\hbar U\left( \mathbf{P},\mathbf{R}\right) \nabla
_{P^{l}}H_{0}\left( \mathbf{P,R}\right) \nabla _{R_{l}}U^{+}\left( \mathbf{%
P,R}\right)  \notag \\
&&+\frac{i}{2}\hbar \nabla _{P^{l}}U\left( \mathbf{P},\mathbf{R}\right)
\nabla _{R_{l}}H_{0}\left( \mathbf{P,R}\right) \nabla _{P^{l}}U^{+}\left(
\mathbf{P,R}\right)  \notag \\
&&+\frac{i}{2}\hbar \nabla _{R_{l}}U\left( \mathbf{P},\mathbf{R}\right)
H_{0}\left( \mathbf{P,R}\right) \nabla _{P^{l}}U^{+}\left( \mathbf{P,R}%
\right)  \notag \\
&&-\frac{i}{2}\hbar \nabla _{P^{l}}U\left( \mathbf{P},\mathbf{R}\right)
H_{0}\left( \mathbf{P,R}\right) \nabla _{R_{l}}U^{+}\left( \mathbf{P,R}%
\right)
\end{eqnarray}
Let us first remark that $H_{0}\left( \mathbf{P,R}\right) $ is already
symmetrized in $\mathbf{R}$ and $\mathbf{P}$. As a consequence $Asym\nabla
_{R_{l}}\nabla _{P^{l}}\left[ H_{0}\left( \mathbf{P,R}\right) \right] =0$.
Actually, remember that the asymmetrization term was the sum of commutators
obtained by pushing the momentum half on the left and half on the right. For
the same reason
\begin{equation}
Asym\nabla _{R_{l}}\nabla _{P^{l}}\left[ U\left( \mathbf{P},\mathbf{R}%
\right) \right] =Asym\nabla _{R_{l}}\nabla _{P^{l}}\left[ U^{+}\left(
\mathbf{P,R}\right) \right] =0
\end{equation}
Now, we introduce the transformed variables and the non projected Berry
Phases at order $\hbar $:
\begin{eqnarray}
\mathbf{r} &=&\left( U\left( \mathbf{P},\mathbf{R}\right) +X\right) \mathbf{R%
}\left( U^{+}\mathbf{\left( \mathbf{P,R}\right) +}X^{+}\right) \simeq
\mathbf{R}+\left[ i\hbar U\left( \mathbf{P},\mathbf{R}\right) \nabla
_{P}U^{+}\left( \mathbf{P},\mathbf{R}\right) \right] =\mathbf{R}+\mathcal{A}%
_{R}  \notag \\
\mathbf{p} &=&U\left( \mathbf{P},\mathbf{R}\right) \mathbf{P}U^{+}\mathbf{%
\left( \mathbf{P,R}\right) }\simeq \mathbf{P}-\left[ i\hbar U\left( \mathbf{P%
},\mathbf{R}\right) \nabla _{R}U^{+}\left( \mathbf{P},\mathbf{R}\right) %
\right] =\mathbf{P}+\mathcal{A}_{P}
\end{eqnarray}
Before going further, we can find some relations on the Berry Phases. Given
that :
\begin{equation}
U\left( \mathbf{P},\mathbf{R}\right) U^{+}\mathbf{\left( \mathbf{P,R}\right)
}=1
\end{equation}
at the zeroth order in $\hbar $ we have the following relations at the first
order in $\hbar $ :
\begin{eqnarray}
\mathcal{A}_{R} &=&i\hbar U\left( \mathbf{P},\mathbf{R}\right) \nabla
_{P}U^{+}\left( \mathbf{P},\mathbf{R}\right) =-i\nabla _{P}U\left( \mathbf{P}%
,\mathbf{R}\right) U^{+}\left( \mathbf{P},\mathbf{R}\right)  \notag \\
\mathcal{A}_{P} &=&-i\hbar U\left( \mathbf{P},\mathbf{R}\right) \nabla
_{R}U^{+}\left( \mathbf{P},\mathbf{R}\right) =i\nabla _{R}U\left( \mathbf{P},%
\mathbf{R}\right) U^{+}\left( \mathbf{P},\mathbf{R}\right)
\end{eqnarray}
Using these results as well as $\nabla _{P^{l}}=\frac{-i}{\hbar }\left[
R_{l},\right] $, $\nabla _{R_{l}}=\frac{i}{\hbar }\left[ P^{l},\right] $ and
inserting the operators $U\left( \mathbf{P},\mathbf{R}\right) $ and $%
U^{+}\left( \mathbf{P},\mathbf{R}\right) $ when needed, we get :
\begin{eqnarray}
&&\frac{i}{2}\hbar Asym\nabla _{R_{l}}\nabla _{P^{l}}\left[ U\left( \mathbf{P%
},\mathbf{R}\right) H_{0}\left( \mathbf{P,R}\right) U^{+}\left( \mathbf{P,R}%
\right) \right]  \notag \\
&=&\frac{-i}{2\hbar }\mathcal{A}_{P^{l}}U\left( \mathbf{P},\mathbf{R}\right) %
\left[ R_{l},H_{0}\left( \mathbf{P,R}\right) \right] U^{+}\left( \mathbf{P,R}%
\right) +\frac{i}{2\hbar }\mathcal{A}_{R_{l}}U\left( \mathbf{P},\mathbf{R}%
\right) \left[ P_{l},H_{0}\left( \mathbf{P,R}\right) \right] U^{+}\left(
\mathbf{P,R}\right)  \notag \\
&&-\frac{i}{2\hbar }U\left( \mathbf{P},\mathbf{R}\right) \left[
R_{l},H_{0}\left( \mathbf{P,R}\right) \right] U^{+}\left( \mathbf{P,R}%
\right) \mathcal{A}_{P^{l}}+\frac{i}{2\hbar }U\left( \mathbf{P},\mathbf{R}%
\right) \left[ P^{l},H_{0}\left( \mathbf{P,R}\right) \right] U^{+}\left(
\mathbf{P,R}\right) \mathcal{A}_{R_{l}}  \notag \\
&&-\frac{i}{2\hbar }\mathcal{A}_{P^{l}}\varepsilon \left( \mathbf{P,R}%
\right) \mathcal{A}_{R_{l}}+\frac{i}{2\hbar }\mathcal{A}_{R_{l}}\varepsilon
\left( \mathbf{P,R}\right) \mathcal{A}_{P^{l}}  \notag \\
&=&\frac{-i}{2\hbar }\mathcal{A}_{P^{l}}\left[ r_{l},\varepsilon \left(
\mathbf{P,R}\right) \right] +\frac{i}{2\hbar }\mathcal{A}_{R_{l}}\left[
p^{l},\varepsilon \left( \mathbf{P,R}\right) \right] -\frac{i}{2\hbar }\left[
r_{l},\varepsilon \left( \mathbf{P,R}\right) \right] \mathcal{A}_{P^{l}}+%
\frac{i}{2\hbar }\left[ p^{l},\varepsilon \left( \mathbf{P,R}\right) \right]
\mathcal{A}_{R_{l}}  \notag \\
&&-\frac{i}{2\hbar }\mathcal{A}_{P^{l}}\varepsilon \left( \mathbf{P,R}%
\right) \mathcal{A}_{R_{l}}+\frac{i}{2\hbar }\mathcal{A}_{R_{l}}\varepsilon
\left( \mathbf{P,R}\right) \mathcal{A}_{P^{l}}  \notag \\
&=&\frac{1}{2}\left[ \mathcal{A}_{R_{l}}\nabla _{R^{l}}\varepsilon \left(
\mathbf{P,R}\right) +\nabla _{R^{l}}\varepsilon \left( \mathbf{P,R}\right)
\mathcal{A}_{R_{l}}\right] +\frac{1}{2}\left[ \mathcal{A}_{P^{l}}\nabla
_{P_{l}}\varepsilon \left( \mathbf{P,R}\right) +\nabla _{P_{l}}\varepsilon
\left( \mathbf{P,R}\right) \mathcal{A}_{P^{l}}\right]  \notag \\
&&-\frac{i}{2\hbar }\mathcal{A}_{P^{l}}\left[ \mathcal{A}_{R_{l}},%
\varepsilon \left( \mathbf{P,R}\right) \right] -\frac{i}{2\hbar }\left[
\mathcal{A}_{R_{l}},\varepsilon \left( \mathbf{P,R}\right) \right] \mathcal{A%
}_{P^{l}}+\frac{i}{2\hbar }\mathcal{A}_{R_{l}}\left[ \mathcal{A}%
_{P^{l}},\varepsilon \left( \mathbf{P,R}\right) \right] +  \notag \\
&&\frac{i}{2\hbar }\left[ \mathcal{A}_{P^{l}},\varepsilon \left( \mathbf{P,R}%
\right) \right] \mathcal{A}_{R_{l}}-\frac{i}{2\hbar }\mathcal{A}%
_{P^{l}}\varepsilon \left( \mathbf{P,R}\right) \mathcal{A}_{R_{l}}+\frac{i}{%
2\hbar }\mathcal{A}_{R_{l}}\varepsilon \left( \mathbf{P,R}\right) \mathcal{A}%
_{P^{l}}
\end{eqnarray}
Rearranging the commutators leads to :
\begin{eqnarray}
&&\frac{i}{2}\hbar Asym\nabla _{R_{l}}\nabla _{P^{l}}\left[ U\left( \mathbf{P%
},\mathbf{R}\right) H_{0}\left( \mathbf{P,R}\right) U^{+}\left( \mathbf{P,R}%
\right) \right]  \notag \\
&=&\frac{1}{2}\left[ \mathcal{A}_{R_{l}}\nabla _{R_{l}}\varepsilon \left(
\mathbf{P,R}\right) +\nabla _{R_{l}}\varepsilon \left( \mathbf{P,R}\right)
\mathcal{A}_{R_{l}}\right] +\frac{1}{2}\left[ \mathcal{A}_{P_{l}}\nabla
_{P_{l}}\varepsilon \left( \mathbf{P,R}\right) +\nabla _{P_{l}}\varepsilon
\left( \mathbf{P,R}\right) \mathcal{A}_{P_{l}}\right]  \notag \\
&&-\frac{i}{2\hbar }\mathcal{A}_{R_{l}}\left[ \varepsilon \left( \mathbf{P,R}%
\right) ,\mathcal{A}_{P_{l}}\right] -\frac{i}{2\hbar }\left[ \varepsilon
\left( \mathbf{P,R}\right) ,\mathcal{A}_{P_{l}}\right] \mathcal{A}_{R_{l}}+%
\frac{i}{2\hbar }\mathcal{A}_{P^{l}}\left[ \varepsilon \left( \mathbf{P,R}%
\right) ,\mathcal{A}_{R_{l}}\right]  \notag \\
&&+\frac{i}{2\hbar }\left[ \varepsilon \left( \mathbf{P,R}\right) ,\mathcal{A%
}_{R_{l}}\right] \mathcal{A}_{P^{l}}-\frac{i}{2\hbar }\mathcal{A}%
_{P^{l}}\varepsilon \left( \mathbf{P,R}\right) \mathcal{A}_{R_{l}}+\frac{i}{%
2\hbar }\mathcal{A}_{R_{l}}\varepsilon \left( \mathbf{P,R}\right) \mathcal{A}%
_{P^{l}}
\end{eqnarray}
rewrite $\frac{-i}{2\hbar }\mathcal{A}_{P^{l}}\varepsilon \left( \mathbf{P,R}%
\right) \mathcal{A}_{R_{l}}+\frac{i}{2\hbar }\mathcal{A}_{R_{l}}\varepsilon
\left( \mathbf{P,R}\right) \mathcal{A}_{P^{l}}$ as $-\frac{i}{2\hbar }%
\mathcal{A}_{P^{l}}\left[ \varepsilon \left( \mathbf{P,R}\right) ,\mathcal{A}%
_{R_{l}}\right] +\frac{i}{2\hbar }\mathcal{A}_{R_{l}}\left[ \varepsilon
\left( \mathbf{P,R}\right) ,\mathcal{A}_{P^{l}}\right] +\frac{i}{2\hbar }%
\left[ \mathcal{A}_{R_{l}},\mathcal{A}_{P^{l}}\right] \varepsilon \left(
\mathbf{P,R}\right) $
\end{quote}

so that :
\begin{eqnarray}
&&\frac{i}{2}\hbar Asym\nabla _{R_{l}}\nabla _{P^{l}}\left[ U\left( \mathbf{P%
},\mathbf{R}\right) H_{0}\left( \mathbf{P,R}\right) U^{+}\left( \mathbf{P,R}%
\right) \right]  \notag \\
&=&\frac{1}{2}\left[ \mathcal{A}_{R_{l}}\nabla _{R^{l}}\varepsilon \left(
\mathbf{P,R}\right) +\nabla _{R^{l}}\varepsilon \left( \mathbf{P,R}\right)
\mathcal{A}_{R_{l}}\right] +\frac{1}{2}\left[ \mathcal{A}_{P^{l}}\nabla
_{P_{l}}\varepsilon \left( \mathbf{P,R}\right) +\nabla _{P_{l}}\varepsilon
\left( \mathbf{P,R}\right) \mathcal{A}_{P^{l}}\right]  \notag \\
&&-\frac{i}{2\hbar }\left[ \varepsilon \left( \mathbf{P,R}\right) ,\mathcal{A%
}_{P^{l}}\right] \mathcal{A}_{R_{l}}+\frac{i}{2\hbar }\left[ \varepsilon
\left( \mathbf{P,R}\right) ,\mathcal{A}_{R_{l}}\right] \mathcal{A}_{P^{l}}+%
\frac{i}{2\hbar }\left[ \mathcal{A}_{R_{l}},\mathcal{A}_{P^{l}}\right]
\varepsilon \left( \mathbf{P,R}\right)
\end{eqnarray}
And we thus have :
\begin{eqnarray}
&&U\left( \mathbf{P},\mathbf{R}\right) H_{0}\left( \mathbf{P,R}\right)
U^{+}\left( \mathbf{P,R}\right)  \notag \\
&=&\varepsilon \left( \mathbf{P,R}\right) +\frac{1}{2}\left[ \mathcal{A}%
_{R_{l}}\nabla _{R^{l}}\varepsilon \left( \mathbf{P,R}\right) +\nabla
_{R^{l}}\varepsilon \left( \mathbf{P,R}\right) \mathcal{A}_{R_{l}}\right] +%
\frac{1}{2}\left[ \mathcal{A}_{P^{l}}\nabla _{P_{l}}\varepsilon \left(
\mathbf{P,R}\right) +\nabla _{P_{l}}\varepsilon \left( \mathbf{P,R}\right)
\mathcal{A}_{P^{l}}\right]  \notag \\
&&-\frac{i}{2\hbar }\left[ \varepsilon \left( \mathbf{P,R}\right) ,\mathcal{A%
}_{P^{l}}\right] \mathcal{A}_{R_{l}}+\frac{i}{2\hbar }\left[ \varepsilon
\left( \mathbf{P,R}\right) ,\mathcal{A}_{R_{l}}\right] \mathcal{A}_{P^{l}}+%
\frac{i}{2\hbar }\left[ \mathcal{A}_{R_{l}},\mathcal{A}_{P^{l}}\right]
\varepsilon \left( \mathbf{P,R}\right)
\end{eqnarray}
as claimed in the text.

\section{Appendix 2. How solid states Physics fits in our Framework.}

In solid state Physics, we assume that the Hamiltonian is invariant through
a discrete group of translations, for example a group of lattice
translations, whose elements have the form
\begin{equation}
T(\mathbf{b)=}\exp \left( -\frac{i}{\hbar }\int_{0}^{\mathbf{b}}A_{i}(%
\mathbf{R}+\mathbf{r)dr}\right) \exp \left( \frac{i}{\hbar }\mathbf{P.b}%
\right)
\end{equation}
where $\mathbf{b}$ is an arbitrary lattice vector. The eigenvalues of this
operator are degenerated and have the form
\begin{equation}
\exp \left( i\mathbf{k}.\mathbf{b}\right)
\end{equation}
where $\mathbf{k}$ belongs to some reduced dual lattice (a fraction of the
dual lattice, i.e. a plaquette in solid state physics). We aim at defining
the generators $\mathbf{K}$ of these transformations as
\begin{equation}
\mathbf{K}.\mathbf{b}=\log \left( T(\mathbf{b)}\right)
\end{equation}
so that we can define
\begin{equation}
K_{i}=\partial _{b_{i}}\mathbf{K}.\mathbf{b}
\end{equation}
However, this logarithm cannot be defined uniquely, and we will build an
explicit choice in the sequel.

To do so, we work with the extended representation, so that the considered
state space is defined, similarly to the Dirac case, by : $%
L^{2}(R^{3})\otimes E$, where $L^{2}(R^{3})$ is seen as the set of functions
of the variable $\mathbf{k}$, running on $R^{3}$. $E$ is a vector space of
infinite size representing the bands. For each value of $\mathbf{k}$, $T(%
\mathbf{b)}$ is diagonal in $E$, with eigenvalues $\exp \left( i\mathbf{k}.%
\mathbf{b}\right) $. We define $K_{i}$ as acting diagonally as the
multiplication by $k_{i}$. In other words, we have defined the momentum
through an extension : $K\equiv K\otimes Id_{E}$. Consequently the position
operator $\mathbf{R}$ is acting as $i\nabla _{\mathbf{k}}$. Therefore the
state space of the Bloch electron is spanned by the basis vector of plane
waves\textbf{\ }$\left| n,\mathbf{k}\right\rangle =\left| \mathbf{k}%
\right\rangle \otimes \left| n\right\rangle $ with $n$ corresponding to a
band index. The state $\left| n\right\rangle $ can be seen as a canonical
base vector\textbf{\ }$\left| n\right\rangle =(0...010...0...)$ (with $1$ at
the $n$th position).

Turning now to the diagonalization process for the Hamiltonian, this last
one can be performed independently for each value of $\mathbf{k}$, since the
Hamiltonian commutes with the translations. We can thus see the Hamiltonian
as a set of square matrices indexed by $\mathbf{k}$, each of them acting on
each copy of $E$. As a consequence the diagonalization matrix is a Block
acting on $E$ for each value of $\mathbf{k}$.

Note that this diagonalization matrix can of course be seen as an operator $%
U(\mathbf{K,R)}$ or, and this is the point of view we adopt here, as a
matrix acting on each copy of $E$, that is, a matrix $U(\mathbf{K)}$, whose
entries depend on $\mathbf{K}$ only. Actually, the dependence in $\mathbf{R}$
appears in the non diagonal elements, and we can discard them if we consider
this ''half matrix, half operator'' version. This mixed representation has
the advantage to do the connection with the Dirac Hamiltonian.

In this set up, $\mathbf{K}$ \ being diagonal and proportional to the
identity, it commutes with every matrix $U(\mathbf{K,R)}$ preserving the
Blocks. When considering the diagonalized Hamiltonian \ $\varepsilon (%
\mathbf{R},\mathbf{K)}$, it can also be seen as a diagonal matrix
(implicitly denoted \ $\varepsilon (\mathbf{K)}$) whose components are
diagonal and denoted $\varepsilon _{n}(\mathbf{K)}$, the $n$ th band energy.
The commutator
\begin{equation}
\left[ \mathbf{R,}\varepsilon (\mathbf{K)}\right] =\nabla \varepsilon (%
\mathbf{K)}
\end{equation}
is again a diagonal matrix whose entries are $\nabla \varepsilon _{n}(%
\mathbf{K)}$ (sketch of proof : $\varepsilon (\mathbf{K)}$ is a series whose
elements are products of powers of $\mathbf{K}$ and $\mathbf{R}$. For each
power of $\mathbf{K}$, the fact that $\ \varepsilon (\mathbf{K)}$ is
diagonal implies that the dependence in $\mathbf{R}$ is a diagonal matrix.
The gradient in $\mathbf{K}$ acting only on the power of $\mathbf{K}$, this
diagonality is preserved.) Turning now to the perturbation $\delta A(\mathbf{%
R})$, let us remark that if the operator $\delta A(\mathbf{R})$ preserves
the bands, all operators $\varepsilon (\mathbf{K)}\delta A(\mathbf{R})$, $%
\left[ \varepsilon (\mathbf{K),}\delta A(\mathbf{R})\right] $ are diagonal,
and given our previous remarks, the same is true for $\nabla \varepsilon (%
\mathbf{K)}\delta A(\mathbf{R})$, $\left[ \nabla \varepsilon (\mathbf{K),}%
\delta A(\mathbf{R})\right] $.

All this remarks that are obviously true for the Dirac case, appear to be
useful in the solid state physics case (application $2$), since it shows
that the electron in an periodic potential fits in our framework. Actually,
we can formally consider the Hamiltonian of such a problem as given by a
matrix depending on the momenta and the coordinates.

\end{document}